\documentclass{article}
\usepackage{setspace}
\usepackage{arxiv}
\usepackage{algorithm}
\usepackage{algpseudocode}
\usepackage[utf8]{inputenc} 
\usepackage[T1]{fontenc}    
\usepackage{hyperref}       
\usepackage{url}            
\usepackage{booktabs}       
\usepackage{amsfonts}       
\usepackage{nicefrac}       
\usepackage{microtype}      
\usepackage{lipsum}
\usepackage{graphicx}
\usepackage{amsmath}
\usepackage[dvipsnames]{xcolor}
\definecolor{darkblue}{RGB}{0,0,139}
\definecolor{midnightblue}{RGB}{25,25,112}
\usepackage{caption}
\usepackage{bm}
\usepackage[font={color=darkblue},figurename=Figure]{caption}
\graphicspath{ {./figures/} }
\usepackage{authblk}
\usepackage[version=4]{mhchem}

\title{A pore-scale model for electrokinetic in situ recovery of copper: the Influence of mineral occurrence, zeta potential, and electric potential}
\author[a,*]{Kunning~Tang}
\author[b,*]{Zhe~Li \thanks{Kunning~Tang and Zhe~Li have same contribution to the publication}}
\author[a]{Ying Da~Wang}
\author[c]{James McClure}
\author[d]{Hongli~Su}
\author[a]{Peyman~Mostaghimi}
\author[a,†]{Ryan T.~Armstrong \thanks{Corresponding author: Ryan T. Armstrong, Telephone: +61 0435500738. Email: ryan.armstrong@unsw.edu.au}}

\affil[a]{School of Minerals and Energy Resources Engineering, University of New South Wales,Sydney, NSW 2052,Australia}
\affil[b]{Research School of Physics, The Australian National University,Canberra, ACT 2601,Australia}
\affil[c]{National Security Institute, Virginia Tech,Blacksburg, VA 24061, USA} 
\affil[d]{Institute of Frontier Materials, Deakin University, Geelong, Victoria 3220, Australia}

\begin{document}
\setcounter{tocdepth}{4}
\setcounter{secnumdepth}{4}
\maketitle

{\centering
\includegraphics[width=0.75\textwidth]{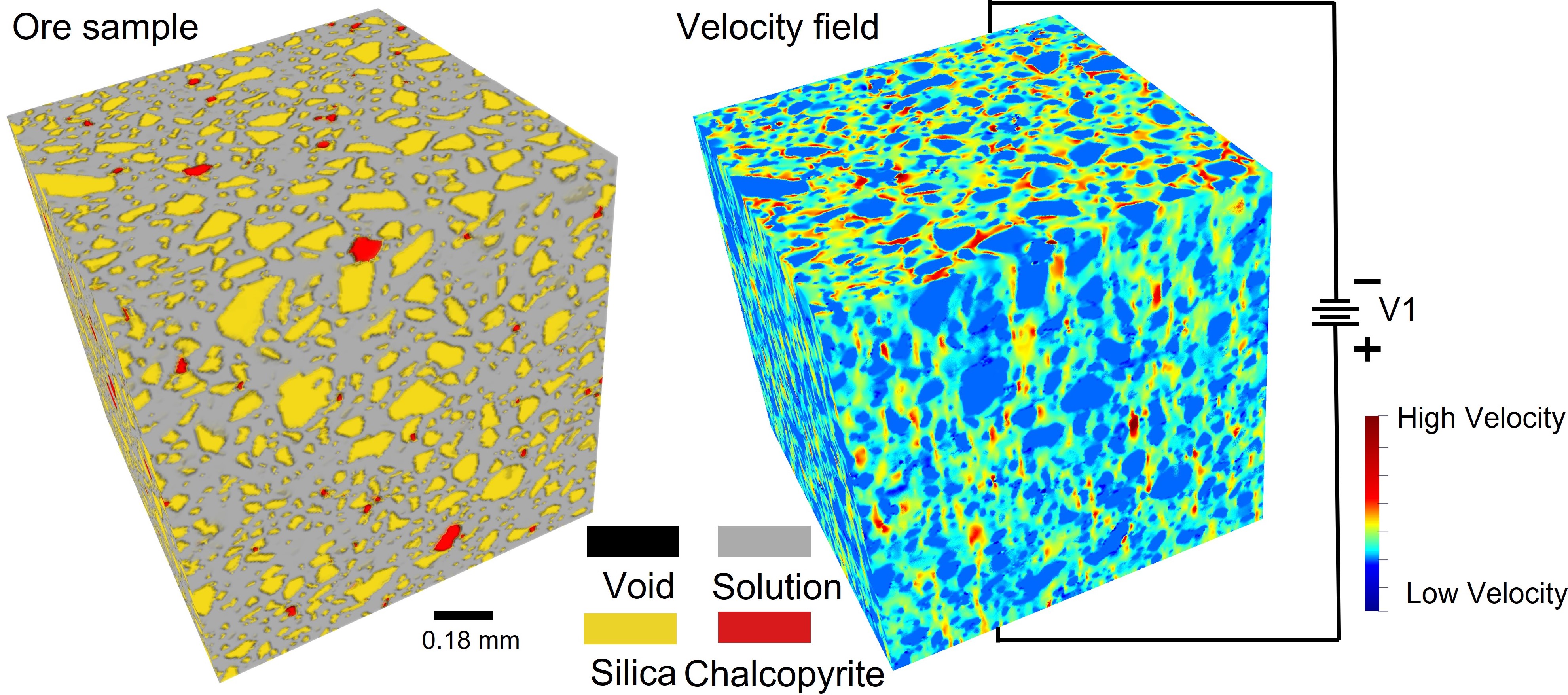}
\par
}

\pagebreak
\begin{abstract}
Electrokinetic in-situ recovery is an alternative to conventional mining, relying on the application of an electric potential to enhance the subsurface flow of ions. Understanding the pore-scale flow and ion transport under electric potential is essential for petrophysical properties estimation and flow behaviour characterization. The governing physics of electrokinetic transport are electromigration and electroosmotic flow, which depend on the electric potential gradient, mineral occurrence, domain morphology (tortuosity and porosity, grain size and distribution, etc.) and electrolyte properties (local pH distribution and lixiviant type and concentration, etc.). Herein, mineral occurrence and its associated zeta potential are investigated for EK transport. The governing model includes three coupled equations: (1) Poisson equation, (2) Nernst--Planck equation, and (3) Navier--Stokes equation. These equations were solved using the lattice Boltzmann method within X-ray computed microtomography images. The proposed model is validated against COMSOL Multiphysics in a 2-dimensional microchannel in terms of fluid flow behavior when electrical double layer is both resolvable and unresolvable. A more complex chalcopyrite-silica system is then obtained by micro-CT scanning to evaluate the model performance. The effects of mineral occurrence, zeta potential, and electric potential on the 3-dimensional chalcopyrite-silica system was evaluated. Although the positive zeta potential of chalcopyrite can induce a flow of ferric ion counter to the direction of electromigration, the net effect is dependent on the occurrence of chalcopyrite. However, the ion flux induced by electromigration was the dominant transport mechanism, whereas advection induced by electroosmosis made a lower contribution. Overall, a pore-scale EK model is proposed for direct simulation on pore-scale images. The proposed model can be coupled with other geochemical models for full physico-chemical transport simulations. Meanwhile, Electrokinetic transport is a promising technique that can be controlled because the dominant ion transport mechanism is electromigration, which depends on the applied external electric potential. 

\end{abstract}


\pagebreak
\section{Introduction}
\label{sec:intro}
\textit{In situ} recovery (ISR), also known as \textit{in situ} leaching, is a technically and commercially feasible method for the extraction of minerals from high-grade ore bodies in a radically different way to conventional mining \cite{paul_economic_1989,seredkin_situ_2016,vargas_situ_2020}. ISR is the circulation of lixiviants through subsurface mineralized ore fractures to dissolve valuable minerals, especially metal minerals, without physically destroying ore formations \cite{ahlness_situ_1983,bates_glossary_1987,council_evolutionary_2002,sinclair_situ_2015}. Compared with conventional mining, ISR has several advantages: (1) it eliminates mining costs, costs for removing ores and fragments to surface dumps, and costs of storage/disposal of tailings \cite{seredkin_situ_2016}; (2) it has lower noise, and greenhouse gas emissions \cite{sinclair_situ_2015}; and (3) it creates a safer working environment for mine workers. Starting in the early 1970s, ISR was developed and applied for uranium extraction from roll-front sandstone deposits, particularly in Kazakhstan and Uzbekistan \cite{seredkin_situ_2016,kuhar_assessment_2018,lagneau_industrial_2019,zhou_uranium_2020}. To date, ISR has been extensively applied in the production of other metals such as copper, gold, and lithium \cite{ogorman_novel_2004,seredkin_situ_2016}. Unlike sandstone deposits that contain many pores and fractures, the host formations for many metallic minerals are mainly low-permeability consolidated hard rocks. Therefore, it is difficult to inject lixiviant into hard rock formations by hydraulic force, which presents challenges for ISR and results in low recovery rates \cite{sinclair_situ_2015}. 

To resolve the issue of the limited permeability of hard rock formations, electrokinetic (EK) ISR (EK-ISR) has been proposed to enhance ion transportation within low-permeability formations via application of an external electric potential \cite{martens_electrokinetic_2018,martens_feasibility_2018,martens_toward_2021}. EK is a mature technique that has been applied in many engineering fields, including soil remediation \cite{virkutyte_electrokinetic_2002}, wastewater treatment \cite{yuan_electrokinetic_2006}, and mine tailing remediation \cite{baek_electrolyte_2009}. For conventional ISR, flow is governed by hydraulic pressure gradients, whereby only a few preferential flow paths can be swept, which makes ISR highly unstable and unpredictable, as these preferential flow paths may host only a small fraction of the total metals \cite{sinclair_situ_2015,martens_toward_2021}. In EK-ISR, electric potential induces a more homogeneous flow through the heterogeneous ore. Therefore, EK-ISR is currently considered a promising method for the recovery of various metals, including gold and copper, from intact hard rocks \cite{martens_electrokinetic_2018,martens_feasibility_2018,martens_toward_2021}. EK-ISR, However, has not yet been applied in any mining project, and thus requires further experimental and numerical work to better understand its efficacy for specific physico-chemical subsurface environments and rock surface properties. Until recently, several researchers experimentally and numerically studied EK transport in porous media and EK-ISR \cite{chowdhury2017electrokinetic,alizadeh2019impact,sprocati2019modeling,tripathi2020electro,martens_toward_2021,gill2021electrokinetic,sprocati2022interplay,karami2022investigation}. Especially, \cite{chowdhury2017electrokinetic,gill2021electrokinetic} studied the electrokinetic permanganate delivery in low-permeable porous media and found that electrokinetics significantly enhanced fluid permanganate delivery. \cite{karami2022investigation} conducted  lab-scale experiments on EK-ISR with different voltage and pressure field and \cite{martens_toward_2021} used COMSOL Multiphysics coupled with Phreeqc to perform a continuum-scale simulation of EK-ISR. However, to our best knowledge, the simulation and experimental EK-ISR study and the comparison between them are still lacking. This digital twin study can be potentially performed with micro-CT imaging and large-domain pore-scale direct simulation \cite{mcclure2014novel,da2019computations,ali2020virtual,xiao2021permeability}. The information extracted from the digital twin study can be used for the upscaling study.

The main transport mechanisms of EK-ISR include (1) electromigration, which involves the displacement of charged species towards an electrode of opposite charge, and (2) electroosmotic flow (EOF), which represents the net movement of fluid flow as a result of the excess charge adhered to the mineral surface \cite{acar1993principles}. The electrical double layer (EDL) plays a key role in EOF. The EDL consists of a charged solid surface and a thin layer of counter ions in an aqueous solution. As counter ions in the EDL move towards the oppositely charged electrode, momentum is transferred to the surrounding fluid molecules, thereby inducing flow \cite{acar1993principles}. The Helmholtz--Smoluchowsky equation (HS) is commonly used to determine EOF in porous media, and its application depends on the thickness of the EDL \cite{acar1993principles, WANG2007264,zhang_electro-osmosis_2017}. Most studies in the literature assume a thin double layer, which means that the thickness of the EDL is considerably smaller than the pore size \cite{WANG2007264, zhang_electro-osmosis_2017}. The counter ions of the EDL screen the wall charge within a region that scales with Debye length \cite{Pennathur2005ElectrokineticTI}. In addition, the thickness of the EDL varies with the electric potential at the particle surface. The zeta potential is a way to characterize the EDL based on the ionic concentration in the EDL and the pH as a result of the protonation/deprotonation reactions that occur at the particle surface \cite{VANE19971,Lima,zhang_electro-osmosis_2017,KHOSO201955,LIU2010542}. Other essential procedures and parameters for EK including geochemical reaction and local pH distribution as well as some larger-scale factors such as tortuosity and porosity were characterized for fluid-solid systems \cite{mattson2002electrokinetic,appelo2007multicomponent,al2008electrokinetic,storey2012effects,zhang_electro-osmosis_2017,sprocati2019modeling,sprocati2020charge,priya2021pore}. Geochemical reactions change ion composition and concentration and therefore, changes the thickness of EDL and zeta potential \cite{pengra1996temperature,al2008electrokinetic}. The dissolution of the mineral during reaction will change the porous structure and flow pattern. These changes due to geochemical reaction influence the electroosmotic permeability and electromigration. Considering EK-ISR, the ionic concentration is high and its effect on the ion transport becomes non-trivial. Meanwhile, the local pH heterogeneity causes the heterogeneity of zeta potential and results in a nonlinear response of the electroosmotic velocity. With a significant change of pH, the zeta potential might be reversed and results in the reversed electroosmotic velocity \cite{zhang_electro-osmosis_2017}. Tortuosity and porosity provide the morphological information of the porous ore and provides the bridge to study the  micro-scale and macro-scale relationship for future upscaling \cite{yeung1994chapters,pengra1995electrokinetic,pengra1999determination}. \cite{pengra1999determination} compared the permeability estimated by EK and generated from flow experiments, the two permeability values are consistent after reconciling through the hydraulic tortuosity. \cite{alizadeh2019impact,sprocati2022interplay} studied the effect of the heterogeneity porosity to the EK and found that the porous heterogeneities play an important role in EK and the coupling to hydraulic process.

Most of the metal ore formation for ISR is composed of a low porosity-permeability hard rock system, with only thin fractures presented. In this case, the thickness of EDL could be comparable to the fracture size. EDL can then be fully described and resolved. If these fractures are connected to the inlet, the velocity front for the lixiviant changes from plug-shape into parabolic-shape \cite{zhang_electro-osmosis_2017}. For a negatively-charged mineral surface in the fracture, the flow velocity towards the cathode is lower near the mineral surface. Therefore, to understand EOF in EK-ISR, the thickness of the EDL and zeta potential must be characterized based on the chemical condition of the ore. EOF in microchannels has been extensively studied because of its significant applications in EK remediation \cite{ACAR1995117,SoilContamination,Pamukcu,Krishna}. However, most studies are based on simplified pore-structure models \cite{WANG2007264,wang_lattice_2006,zhang_electro-osmosis_2017,Alizadeth}. \cite{wang_lattice_2006} developed a numerical method to simulate electroosmotic flow in a 2D microchannel, which consisted of the combination of nonlinear Poisson equation for the electric potential with lattice Boltzmann method (LBM) for fluid flow \cite{wang_lattice_2006, yoshida_coupled_2014,basu_fully_2020,li_lbm_2021}. Most EK studies using LBM are based on simultaneously solving the Poisson, Nernst--Planck, and Navier--Stokes equations in a self-consistent scheme. \cite{wang_lattice_2006} studied the effect of electrically-driven and pressure-driven flows on flow velocity, electro-viscous effect, and electroosmosis in homogeneous microchannels. The LBM has also been used to study EOF in a heterogeneous pore structure reconstructed using random porous structures \cite{zhang_electro-osmosis_2017}. They studied the effects of the heterogeneous zeta potential at the mineral-liquid interface at different pH values and reported that for a small electric potential strength, the effect of the electrical force on the distribution of pH causes a nonlinear response in the electroosmotic velocity. Such studies are examples of the successful application of numerical schemes based on the LBM to analyzing EK flows. However, these studies mainly focused on the fundamental physical perspective of EOF and used manually generated porous media as the study domain. At the same time, the effect of EOF on the flow behavior in complex porous media has not yet been investigated. 

Herein, we developed a EK model for simulation the physical transport of fluid and ions for EK using the lattice Boltzmann--Poisson method (LBPM) based on the fundamental principles of surface chemistry and EK transport. The governing model includes three coupled equations: (1) Poisson equation, (2) Nernst--Planck equation, and (3) Navier--Stokes equation. In this study, we describe the main workflow of our model and validation in terms of electroosmosis. Meanwhile, a more complex chalcopyrite-silica system is investigated and the efficacy of ion transport under various EK conditions and mineral distributions is evaluated. Specifically, the EK model based on LBPM was first validated against COMSOL Multiphysics in terms of EOF in cases where the EDL is both resolvable and unresolvable in a 2D microchannel. After validation, a chalcopyrite-silica system was created by mixing silica and chalcopyrite powders. The synthetic ore was imaged using high-resolution X-ray micro-computed tomography (micro-CT), which allowed the visualization of the 3D structure and mineral distribution within the system and subsequent direct simulation with LBPM. Overall, the proposed model was designed to investigate the transport of fluid/ions to the target metals under electric potential at the pore-scale. For future work, considering the importance of surface potential in this study, a surface complexation model is planned to be introduced to characterize zeta potential at liquid-solid interfaces. The proposed model can be coupled with a geochemical solver such as PhreeqcRM \cite{parkhurst1995user,parkhurst1999user,parkhurst2015phreeqcrm} for the full characterization of physical-chemical process in the EK-ISR.

\section{Methods and Materials}
\label{sec:Methods and ore sample description}

\subsection{Numerical Methods}
\label{sec:Numerical}

The governing model of the EK flow includes three coupled equations: Poisson equation for the electric potential, Nernst--Planck equation for ion transport driven by chemical and electric potentials, and Navier--Stokes equation for the flow of an electrolyte solution carrying ions.

\subsubsection{Mathematical Models}
\label{sec:Mathematical}
The flow of the electrolyte solution is governed by the incompressible conservation of mass and Navier--Stokes equation:

\begin{equation}\label{eq:NavierStokesEqs}
    \begin{aligned}
        \nabla \cdot \bm{u} &= 0, \\
        \rho_0 \frac{\partial \bm{u}}{\partial t} + \rho_0 \bm{u} \cdot \nabla \bm{u} &= -\nabla p + \mu \nabla^2 \bm{u} + \bm{F},
    \end{aligned}
\end{equation}

where $\bm{u}$ is the fluid velocity vector, $\rho_0$ is the fluid density, $p$ is the fluid pressure, $\mu$ is the dynamic viscosity, and $\bm{F}$ is the body force, which in this study is primarily caused by an external electric potential. The Navier--Stokes equations were solved in pore spaces, whereas a standard nonslip boundary condition was applied to the solid spaces. In the case where the thickness of the EDL was much smaller than the characteristic length of the simulation, that is, below the resolution of an input image, an electroosmotic velocity boundary condition was introduced in the EDL which excluded the detailed flow field between the solid and slipping plane, and analytically calculated the velocity at the solid according to the local zeta potential of the solid surface. Herein, we adopted the commonly used HS equation

\begin{equation}\label{eq:HSequation}
    \bm{u} = - \frac{\epsilon \zeta}{\mu} \nabla_T \psi,\;\; \text{for}\; \bm{x} \in \partial \Omega,
\end{equation}

where $\Omega$ denotes the fluid domain, $\psi$ is the electric potential within the electrolyte, $\zeta$ is the local zeta potential of the solid surface, $\epsilon$ is the permittivity of the electrolyte solution, and $\nabla_T$ is the tangential part of the gradient operator, perpendicular to the solid surface orientation. When the electroosmotic velocity boundary condition is applied as the driving force of the flow, the electric body force in Eq.\ref{eq:NavierStokesEqs} is set to zero, because the EDL is below resolution and the bulk fluid is considered electrically neutral \cite{zhang_electro-osmosis_2017}. 

Ion transport is governed by the Nernst--Planck equation, which incorporates electrochemical migration as an extra drift term into the mass flux:

\begin{equation}\label{eq:NernstPlanckEq}
    \frac{\partial C_i}{\partial t} + \nabla \cdot \left[\left(\bm{u} - \frac{z_i D_i}{V_T} \nabla \psi  \right) C_i \right] = D_i \nabla^2 C_i, 
\end{equation}
where $C_i$ is the concentration of the $i$th ion, $z_i$ is the ion algebraic valency, and $V_T=k_B T/e$ is the thermal voltage, where $k_B$ is the Boltzmann constant and $e$ is the electron charge. The ion mass flux is affected by three factors: the first and second terms on the left-hand side of Eq.\ref{eq:NernstPlanckEq}, which are the convection and electrochemical migration, respectively; the term on the right-hand side is the diffusion where $D_i$ is the diffusivity of the $i$th ion.

A non-flux boundary condition at the fluid--solid interface was applied to the ions:

\begin{equation}\label{eq:non-flux_BC_ion}
    \bm{n}_s \cdot \bm{J}_i = 0,\;\; \text{for}\; \bm{x} \in \partial \Omega,
\end{equation}
where $\bm{n}_s$ is the unit normal vector of the solid surface, and $\bm{J}_i = -D_i \nabla C_i + (\bm{u}-z_i D_i/V_T \nabla \psi) C_i$ is the flux of the $i$th ion. 

The electric potential of the distribution of ions was solved by the Poisson equation:

\begin{equation}
    \nabla^2 \psi = - \frac{\rho_e}{\varepsilon_r \varepsilon_0},
\end{equation}
where $\varepsilon_0$ is the permittivity of vacuum, and $\varepsilon_r$ is the dielectric constant of the electrolyte solution. The net charge density $\rho_e$ (C/m$^3$) is related to the ion concentration as follows:

\begin{equation}
    \rho_e = \sum_i F z_i C_i,
\end{equation}
where the sum runs over all ionic species and $F$ is Faraday's constant given by $F=e N_A$, where $N_A$ is Avogadro's number. The force on the body owing to the net charge density in the Navier—Stokes equation is given by:

\begin{equation}
    \bm{F}_e = \rho_e \bm{E} =- \rho_e \nabla \psi.
\end{equation}

\begin{figure}[H]
  \centering
    \includegraphics[width=0.6\textwidth]{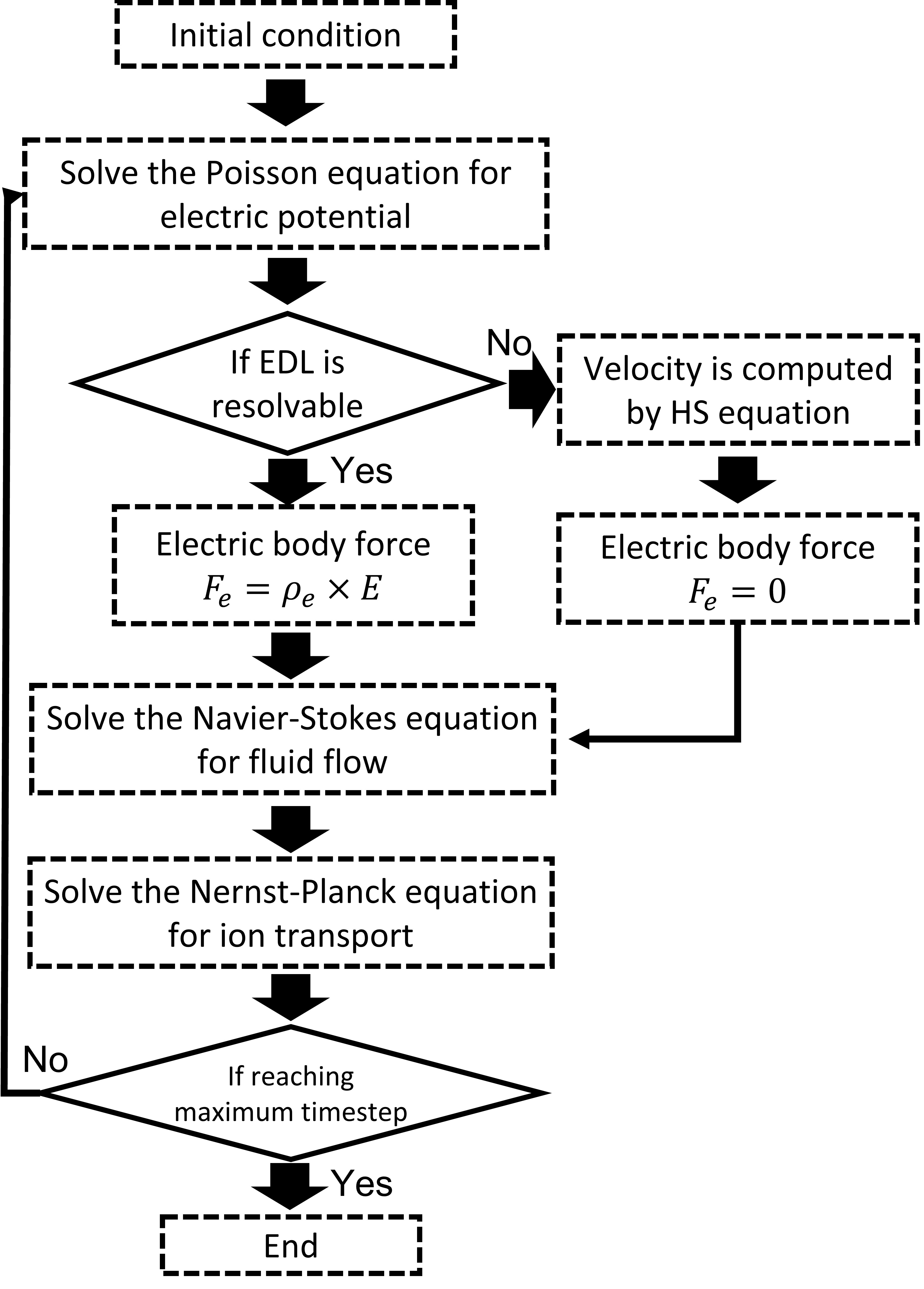}
    \caption{Flow chart for the proposed EK model. Debye length was calculated and compared with voxel length to check if the EDL is resolvable.}
    \label{fig:Modelflow}
\end{figure}

The fluid--solid boundary condition for the electric potential is typically specified in two forms: 1) the surface charge density $\sigma_e$ and 2) surface potential $\psi_s$ at the solid surface (when the EDL is unresolved, the surface potential is equivalent to the zeta potential). The former is a Neumann-type boundary condition given by

\begin{equation}
    \bm{n}_s \cdot \nabla \psi = - \frac{\sigma_e}{\varepsilon_r \varepsilon_0}, \;\; \text{for}\; \bm{x} \in \partial \Omega,
\end{equation}
whereas the latter is a Dirichlet-type boundary given by

\begin{equation}
    \psi(\bm{x}) = \psi_s,\;\; \text{for}\; \bm{x} \in \partial \Omega,
\end{equation}
where $\psi_s$ is the user-specified electric potential of the solid surface. A detailed flow chart of how these equations were solved is shown in Fig. \ref{fig:Modelflow}.

\subsubsection{Lattice Boltzmann Methods}
\label{sec:lbm}

To solve the coupled transport and electric equations (mentioned above) in porous media, we adopted the commonly used LBM because of its inherent scalability of parallel computation and efficient handling of complex boundary conditions. Several coupled lattice Boltzmann (LB) frameworks dedicated to EK flow have been developed and studied over the past decade \cite{wang_lattice_2006, yoshida_coupled_2014, zhang_electro-osmosis_2017, basu_fully_2020}. Herein, we adopted the method proposed by \cite{wang_modeling_2010}, which was modified by \cite{yoshida_coupled_2014,zhang_electro-osmosis_2017}, to incorporate an electroosmotic velocity boundary condition. 

The LB method naturally suits parabolic partial differential equations. The Poisson equation, however, is elliptical, and thus an artificial time-dependent term is usually added so that the LBM yields a steady-state solution of the `transient' Poisson equation of the following form:

\begin{equation}\label{eq:Transient_Poisson_Eq}
    \frac{\partial \psi}{\partial t} = \nabla^2 \psi + \frac{\rho_e}{\varepsilon_r \varepsilon_0}.
\end{equation}
We deployed the D3Q7 lattice to solve the Poisson equation. The corresponding LB evolution equation for the distribution function $h_q$ of the electric potential $\psi$ is given by

\begin{equation}\label{eq:LB_Poisson_Eq}
h_q(\bm{x}+\bm{\xi_q} \triangle x, \tilde{t} + 1) -
h_q(\bm{x}, \tilde{t}) = -\frac{1}{\tau_\psi} \Big[h_q(\bm{x}, \tilde{t}) -h_q^{eq}(\bm{x}, \tilde{t}) \Big] 
+\omega_q \frac{\rho_{e}} {\varepsilon_r \varepsilon_0},
\end{equation}

with
\begin{equation}
    \psi = \sum_{q=0}^{6} h_q,
\end{equation}
and the equilibrium distribution is

\begin{equation}
    h^{eq}_q = \omega_q \psi,
\end{equation}
where  $\bm{\xi}_q$ and $\omega_q$ are the D3Q7 lattice velocity vector and weighting coefficient, respectively, with $\omega_0=1/4$ and $\omega_{1-6}=1/8$; and $\triangle x$ is the spatial resolution of the simulation domain. It should be noted that the time in the LB Poisson equation is denoted as $\tilde{t}$; thus, it should be differentiated from the LB time $t$ in the Navier--Stokes and Nernst--Planck equations to be covered later because only the steady-state solution of Eq.\ref{eq:Transient_Poisson_Eq} is of interest. Within each main evolution step ( Fig. \ref{fig:Modelflow} for flow chart), Eq.\ref{eq:LB_Poisson_Eq} is executed iteratively until the standard mean squared error over certain amount of timestep, $\tilde{t}'$, is smaller than the user specified tolerance:

\begin{equation}
    \frac{1}{N} \sum_{\bm{x}} \left[\psi(\bm{x},\tilde{t}) - \psi(\bm{x},\tilde{t}-\tilde{t}') \right]^2  < \epsilon_{\text{err}},
\end{equation}
where $N$ is the total number of fluid nodes, and $\epsilon_{\text{err}}$ is the prescribed tolerance. 

The LB relaxation time parameter $\tau_{\psi}$ in Eq.\ref{eq:LB_Poisson_Eq} is given by:

\begin{equation}
    \tau_{\psi} = \frac{1}{2}+ \frac{1}{c_s^2},
\end{equation}
where $c_s^2$ is the LB speed of sound; for the D3Q7 lattice $c_s^2=1/4$. 

For the boundary condition, we adopted the formulation developed by \cite{yoshida_coupled_2014}, which is a completely localized scheme that is more suitable for complex porous media. For the Neumann-type boundary condition, where a surface charge density $\sigma_e$ is specified, after the LB collision, the normal streaming step is replaced by the following equation:

\begin{equation}
    h_q(\bm{x}+\bm{\xi_q} \triangle x, \tilde{t} + 1) = h'_{\overline{q}}(\bm{x}+\bm{\xi_q} \triangle x, \tilde{t}) + \sigma_e/ \left(\varepsilon_r \varepsilon_0 \right),
\end{equation}
where $h'_q(\bm{x},t)$ is the post-collision distribution, and the index $\overline{q}$ indicates the direction opposite to $q$. The Dirichlet-type boundary condition, where the surface potential is specified, is given by:

\begin{equation}
    h_q(\bm{x}+\bm{\xi_q} \triangle x, \tilde{t} + 1) = -h'_{\overline{q}}(\bm{x}+\bm{\xi_q} \triangle x, \tilde{t}) + c_s^2 \psi_s.
\end{equation}

Incidentally, the electric field $\bm{E}$ is given by the gradient of the electric potential, that is, $\bm{E} = -\nabla \psi$. According to \cite{yoshida_coupled_2014}, the gradient can be calculated locally as

\begin{equation}
    E_{\alpha} = \frac{1}{\tau_{\psi} c_s^2 \triangle x} \sum_q \xi_{q,\alpha} h_q,
\end{equation}
where the index $\alpha$ denotes Cartesian coordinates. 

For ion transport, the LB evolution equation was also solved for the D3Q7 lattice and is given by 

\begin{equation}
g_q(\bm{x}+\bm{\xi_q} \triangle x, t +\triangle t_{D_i}) -
g_q(\bm{x}, t) = -\frac{1}{\tau_{D_i}} \left[g_q(\bm{x}, t) -g_q^{eq}(\bm{u}, \bm{u}_{EP,i}, t)\right],
\end{equation}
for the ion distribution function $g_q$. The equilibrium distribution function is given by

\begin{equation}
    g_q^{eq} = \omega_q C_i \left[1+ \frac{\triangle t_{D_i}}{\triangle x} \frac{\bm{\xi}_q \cdot \left( \bm{u} + \bm{u}_{EP,i} \right)}{c_s^2} \right],
\end{equation}
where $\triangle t_{D_i}$ is the time resolution; the method of determining $\triangle t_{D_i}$ is covered further on; $C_i=\sum_q g_q$ is the ion concentration of $i$th species; and $\bm{u}_{EP,i}$ is the electrophoretic velocity of the $i$th ion in response to applied electric potential. The electrophoretic velocity is given by

\begin{equation}
    \bm{u}_{EP,i} = \frac{z_i D_{i,\text{LB}}}{V_T} \bm{E} =-\frac{z_i D_{i,\text{LB}}}{V_T} \nabla \psi,
\end{equation}
where $D_{i,\text{LB}}$ is the diffusivity of the $i$th ion species in the LB unit; it is related to the relaxation parameter $\tau_{D_i}$ as follows:

\begin{equation}\label{eq:LB_ion_tau}
    \tau_{D_i} = \frac{1}{2} + \frac{D_{i,\text{LB}}}{c_s^2}.
\end{equation}

For the Dirichlet-type boundary condition, that is, if the surface ion concentration $C_s$ is specified, the normal streaming step after LB collision is replaced by

\begin{equation}
    g_q(\bm{x}+\bm{\xi_q} \triangle x, t + \triangle t_{D_i}) = -g'_{\overline{q}}(\bm{x}+\bm{\xi_q} \triangle x, t) + c_s^2 C_0.
\end{equation}
Here, $g'_q$ denotes the post-collision distribution. For the non-flux boundary condition in Eq.\ref{eq:non-flux_BC_ion}, it has been proved in \cite{yoshida_coupled_2014} that it is equivalent to the half-way bounce-back boundary condition widely used in the LB method, which ensures no ion flux across the solid boundary:

\begin{equation}
    g_q(\bm{x}+\bm{\xi_q} \triangle x, t + \triangle t_{D_i}) = g'_{\overline{q}}(\bm{x}+\bm{\xi_q} \triangle x, t).
\end{equation}

Regarding the LB Navier--Stokes solver for the electrolyte solution, because it has been extensively studied and used in numerous publications, the details of the formulation are not repeated here. Therefore, we implemented the formulation by \cite{McClure2014,McClureLi_AdpLBM_2021,McClureLi_LBPM_2021}, where a multi-relaxation LB method is deployed, and incorporated the slipping velocity boundary condition proposed by \cite{ladd_1994,zhang_electro-osmosis_2017} into model cases in which the EDL is not resolved. 

When solving a multi-physics problem where each transport equation has its own time scale and internal LB timestep, it is important to ensure that all of the coupled equations are synchronized in terms of a physical time scale. In other words, the relationship

\begin{equation}
    N_{t_u} \triangle t_u = N_{t_{D_1}} \triangle t_{D_1} = N_{t_{D_2}} \triangle t_{D_2} = ... = N_{t_{D_n}} \triangle t_{D_n}
\end{equation}
 Must be maintained, where $N_{t_u}$ and $N_{t_{D_i}}$ are the internal LB timestep relative to the main timestep for the Navier--Stokes and ion transport solvers, respectively; $\triangle t_u$ and $\triangle {t_{D_i}}$ are the time conversion factors (for example, unit of [s/l.t.], where l.t. denotes the LB timestep)  for the fluid and $i$th ion species, respectively. 
 

The time conversion factor $\triangle t_u$ for fluid flow is determined by the following relation:

\begin{equation}
    \nu_{\text{phys}} = \frac{\triangle x^2}{\triangle t_u} \nu_{\text{LB}},
\end{equation}
where $\nu_{\text{phys}}$ and $\nu_{\text{LB}}$ are the fluid kinematic viscosities in the physical and LB units, respectively. Notably, $\nu_{\text{LB}}$ is linked to the Navier--Stokes LB relaxation time by $\nu_{\text{LB}} = (\tau_u-0.5)/3$, where $\tau_u$ is usually taken between 0.5 and 2 for numerical stability.  

The time conversion factor $\triangle t_{D_i}$ for ion transport is determined using the following relation:

\begin{equation}
    D_{i,\text{phys}} = \frac{\triangle x^2}{\triangle t_{D_i}} D_{i,\text{LB}},
\end{equation}
where $D_{i,\text{phys}}$ and $D_{i,\text{LB}}$ are the diffusivities of the $i$th ion in physical and LB units, respectively. Note that $D_{i,\text{LB}}$ is related to the LB relaxation time through Eq.\ref{eq:LB_ion_tau}, where $\tau_{D_i}$ is set to 1.0 for numerical stability. 

In summary, using the image resolution $\triangle x$, the input physical parameters ($\nu_{\text{phys}}$, $D_{i,\text{phys}}$) and user specified LB relaxation time ($\tau_u$ and $\tau_{D_i}$), the time conversion factor for each solver was determined; this step was performed using a multi-physics controller in LBPM. The internal LB timestep of each solver was also subsequently determined based on $\triangle t_u$ and $\triangle t_{D_i}$. For example, if $\triangle t_u$ is the largest among the sets $\{\triangle t_u, \triangle t_{D_1}, \triangle t_{D_2},...,\triangle t_{D_n} \}$, then $N_{t_u}$ is set to 1 and $N_{t_{D_i}}$ can be determined using $N_{t_{D_i}}=\triangle t_u/\triangle t_{D_i}$, rounded up to the nearest integer. The open-source code for our EK model is accessible on GitHub (https://github.com/OPM/LBPM).

Our EK model was first benchmarked with COMSOL Multiphysics for EOF under both conditions, where the EDL is resolvable and unresolvable in a two-dimensional microchannel. The validation results can be found in Supplementary Materials Section 3. After validation, we use a simple EK model to evaluate the benefits of EK transport compared to pressure-driven flow in a simple heterogeneous system. EK simulations were then performed on the chalcopyrite-silca system to evaluate the effect of the zeta potential and electric potential on the feasibility of EK transport for EK-ISR. All simulation parameters are listed in Table S1 in Supplementary Materials Section 3. Simulations were performed on a local workstation with 64-core CPU, 24 GB of GPU memory and 256 GB of RAM. Simulations were computed on the GPU with a much faster computational speed than the CPU.

\subsection{Micro-CT imaging and Image Processing}
\label{sec:IronOre}
\begin{figure}[htp!]
  \centering
    \includegraphics[width=\textwidth]{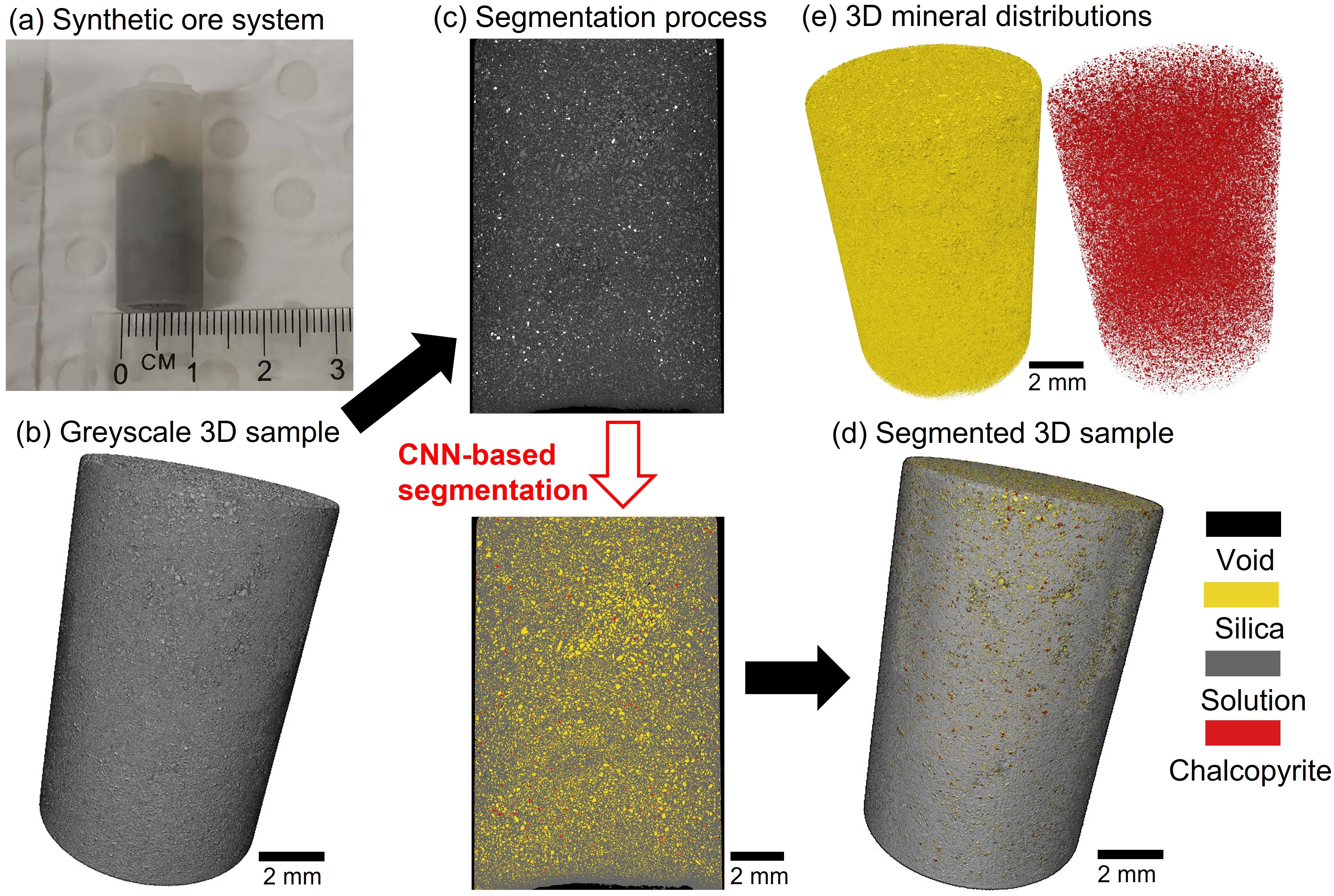}
    \caption{(a) Micro-CT image of the wet chalcopyrite-silica system. The system is a mixture of silica and chalcopyrite powders saturated with sodium chloride solution. (b) Segmentation process identifying 4 phases, including void, solution, silica, and chalcopyrite. Solution used herein is the 0.1 M NaCl. Multi-phase segmentation is based on CNN. (c) Segmented 3D synthetic ore system. (d) 3D distributions of silica and chalcopyrite in the system.}
    \label{fig:Oredata}
\end{figure}

A chalcopyrite-silica system, which is a mixture of a Cu mineral (chalcopyrite) and gangue mineral (silica), was prepared to obtain a copper-rich porous system. This chalcopyrite-silica system was imaged using micro-CT to generate a digital 3D model for simulation. The chalcopyrite powder was obtained from Kremer Pigmente (Germany) with a particle size of approximately 80 $\mu$m. The mineral and elemental contents of the chalcopyrite powder were quantified using X-ray diffraction (XRD) analysis and X-ray fluorescence (XRF), and the results are shown in Table \ref{tab:xrdxrf}. The XRF results show that the major elements are Fe$^{+3}$ and Cu$^{+2}$ which are the main elements in chalcopyrite (CuFeS$_2$). It can be further confirmed from XRD that the powder contained over 72\% chalcopyrite. To prepare the synthetic ore system, SiO$_2$ (2.5 g) and copper ore (40 mg) were mixed in NaCl solution (1.37 mL, 0.1 M) in a beaker (50 mL). The mixture was stirred for 5 min to ensure complete saturation of the system. The saturated mixture was moved to a container (height: 3 cm; inner diameter: 1cm; outer diameter: 1.2 cm) for micro-CT scanning, as shown in Fig. \ref{fig:Oredata} (a). Cotton cloth was used to cover the top of the chalcopyrite-silica system, ensuring minimal powder movement during the micro-CT scanning. The micro-CT image of the chalcopyrite-silica system is shown in Fig. \ref{fig:Oredata} (b). The voxel size was $2013 \times 2013 \times 2970$ with a resolution of 5.4 $\mu$m.

\renewcommand{\arraystretch}{1.6}
\begin{table}[htp!]
\centering
\caption{XRF and XRD results of the chalcopyrite powder. For XRF, powder was preoxidized before being fused into glass bead. Therefore, the XRF results show the element oxide weight percent. The XRD pattern shows the minerals occurring in the powder. Both results confirm that the most abundant mineral in the powder is chalcopyrite.}
\begin{tabular}{lllllllllllll}
\hline
\textbf{XRF} & Na$_{2}$O & MgO    & Al$_{2}$O$_{3}$        & SiO$_{2}$          & SO$_{3}$           & K$_{2}$O           & CaO           & TiO$_{2}$           & Fe$_{2}$O$_{3}$           & CuO           & ZnO         & Loss on ignition \\ \hline
wt.\%                          & 1.09          & 0.22   & 0.68         & 3.71          & 0.82          & 0.22          & 1.04          & 0.05           & 38.96           & 34.48         & 1.34        & 19.07                         \\ \hline
\textbf{XRD}                            & Quartz        & Pyrite & Chalcopyrite & \multicolumn{2}{l}{Covellite} & \multicolumn{2}{l}{Muscovite} & \multicolumn{2}{l}{Anorthoclase} & \multicolumn{2}{l}{Azurite} & Sphalerite                    \\ \hline
wt.\%                          & 5.4           & 8.4    & 72.1         & \multicolumn{2}{l}{0.1}       & \multicolumn{2}{l}{2.6}       & \multicolumn{2}{l}{3.1}          & \multicolumn{2}{l}{7.3}     & 1                            \\ \hline
\end{tabular}
\label{tab:xrdxrf}
\end{table}

Identifying the mineral composition and distribution in the synthetic ore system was a pre-process for EK modeling using LBPM. Therefore, the first step was to perform multiphase segmentation of the synthetic ore system. The synthetic ore system was a mixture of silica, chalcopyrite, NaCl solution, and unsaturated voids. To segment these four phases, we used trainable WEKA segmentation \cite{arganda-carreras_trainable_2017} to generate a training dataset of registered images that were then fed to a U-ResNet convolutional neural network(CNN) \cite{tang_3DMLA_2021,wang_deep_2021,tang_deep_2022,tang_generalizable_2022}. The workflow is outlined as follows.
\begin{enumerate}
\item A 2D grayscale image was selected, and its pixels were manually clustered into the 4 phases as training data for the WEKA segmentation. 
\item After training, the WEKA segmentation generated a 2D segmented slice of the input 2D grayscale image which works as ground truth for CNN training.
\item Once the CNN was trained, it could segment the entire 3D image of the synthetic ore system. The detailed U-ResNet architecture and training schedule can be found in Supplementary Materials Section 1.
\end{enumerate}

Figs. \ref{fig:Oredata} (c--d) show the 3D segmentation result of the 4-phase synthetic ore system, and Fig. \ref{fig:Oredata} (e) shows the 3D distribution of silica and chalcopyrite, which are homogeneously distributed throughout the system.

\section{Results and Discussion}
\label{sec:Results and Discussion}

\subsection{Electrokinetic Model Validation}
\label{sec:EK}
A detailed validation of the EK model built in LBPM is presented in this section. The EOF effect is validated against COMSOL Multiphysics in a 2D microchannel. The constant physical parameters used for validation are listed in Table S\ref{tab:parameter}. After validation, single phase flow behavior under only hydraulic pressure and electric potential is compared in the multi microchannels system. 

\renewcommand{\arraystretch}{1.6}
\begin{table}[htp!]
\centering
\caption{Physical parameters used in the EK numerical simulation.}
\begin{tabular}{lll}
\hline
\textbf{Physical parameter}             & \textbf{Symbol}         & \textbf{value} \\ \hline
Density of electrolyte solution         &  $\rho$                       & 998.2 $kg/m^3$    \\
Temperature                             &    $T$                     & 293.15 $K$       \\
Kinematic viscosity of electrolyte solution  &  $\upsilon$                        & $1.003\times10^{-6} Pa s$  \\
Fluid dielectric constant               &    $\epsilon_{R}$                     & 78.5           \\
Permittivity of vacuum                  &    $\epsilon_{0}$          & $8.85\times10^{-12} F/m$               \\
Diffusivity of H$^+$                      &   $D_{H^+}$                      & $9.3\times10^{-9} m^2/s$      \\
Diffusivity of OH$^+$                       &  $D_{OH^+}$                        & $5.3\times10^{-9} m^2/s$      \\
Diffusivity of Na$^+$                       &  $D_{Na^+}$                        & $1.3\times10^{-9} m^2/s$      \\
Diffusivity of Cl$^+$                       &  $D_{Cl^+}$                        & $2.0\times10^{-9} m^2/s$      \\
Diffusivity of Fe$^{3+}$                       &  $D_{Fe^{3+}}$                        & $6\times10^{-10} m^2/s$      \\ \hline
\end{tabular}
\label{tab:parameter}
\end{table}

\subsubsection{Validation of EOF for the EK model}
\label{sec:EOF}

\begin{figure}[htp!]
  \centering
    \includegraphics[width=\textwidth]{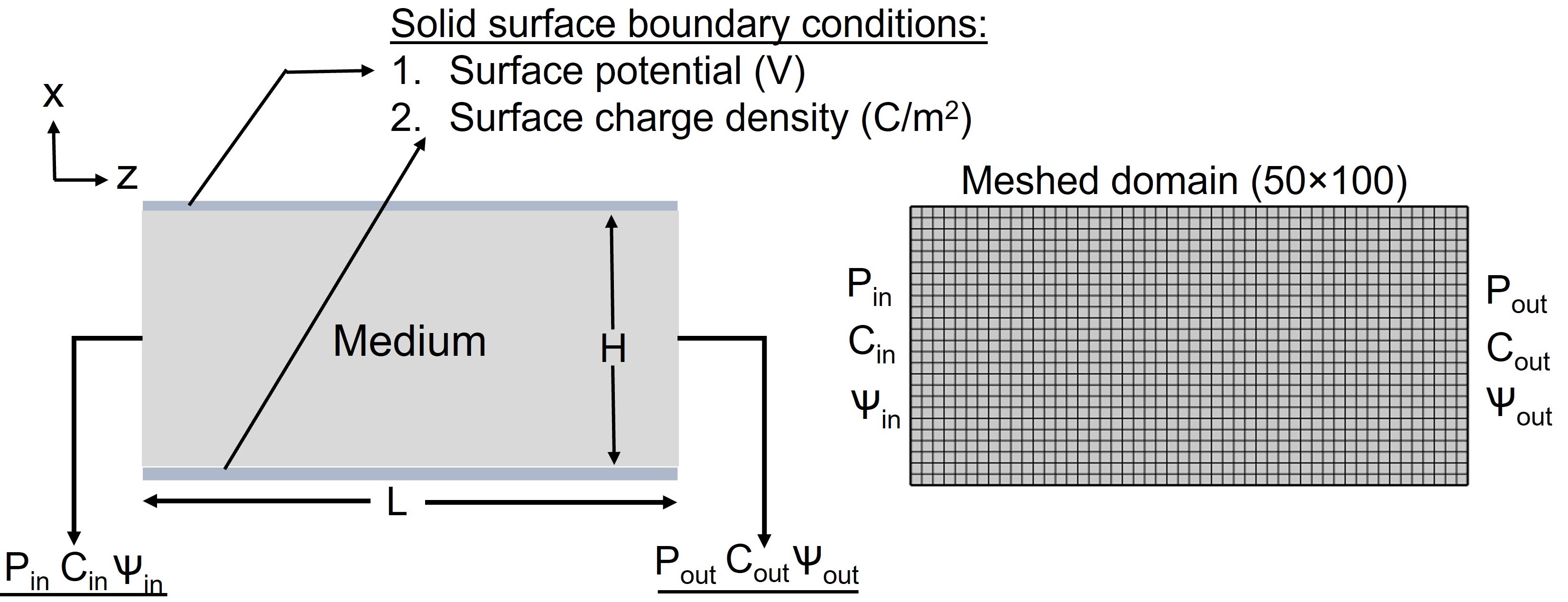}
    \caption{A visualization of the micro-channel and the meshed domain used for validation against COMSOL has 50 pixels in X direction and 100 pixels in Z direction. The physical size is the same as the input for LBPM.}
    \label{fig:Validation}
\end{figure}

\begin{figure}[htp!]
  \centering
    \includegraphics[width=\textwidth]{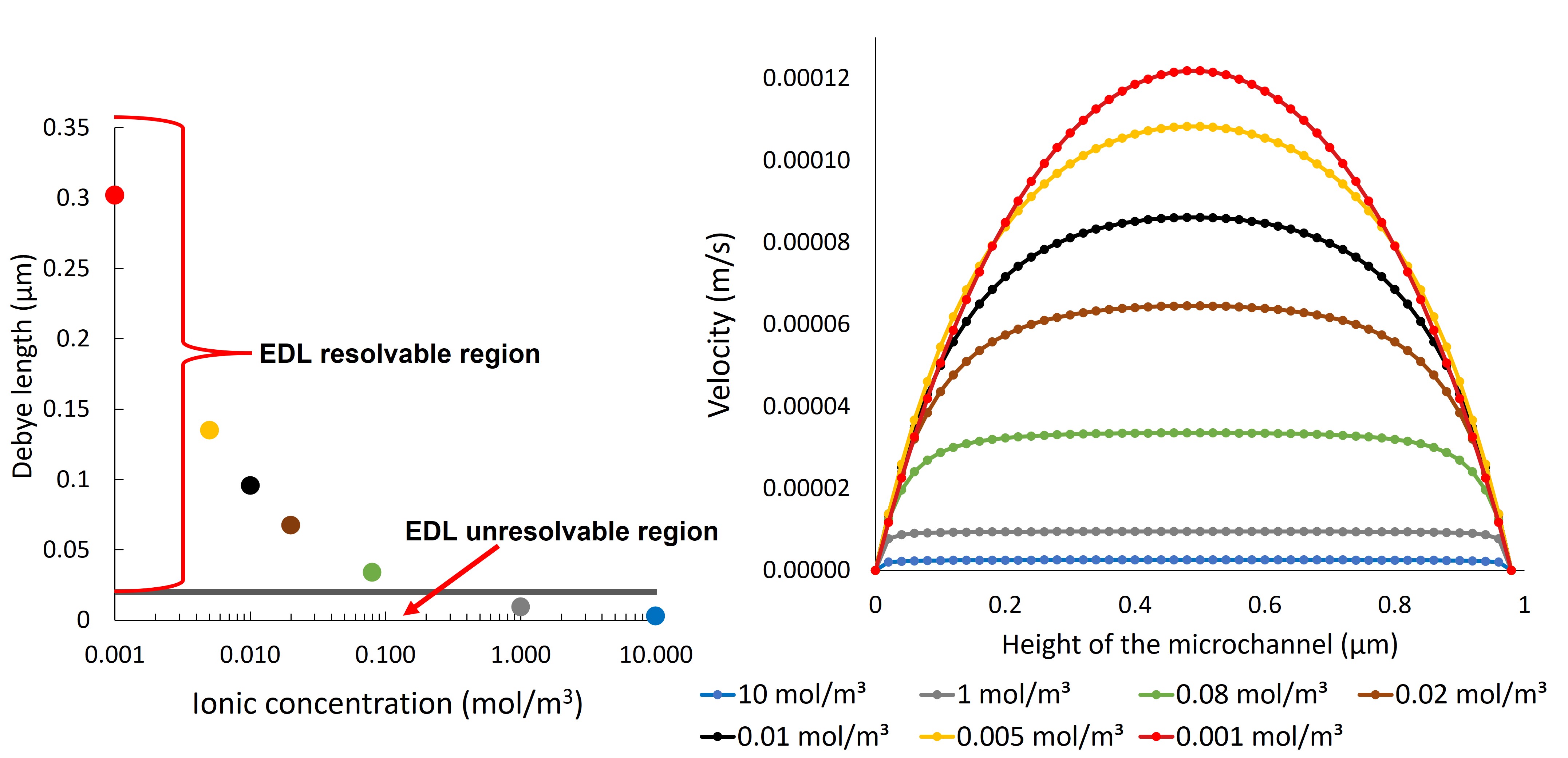}
    \caption{The effect of ionic concentration on the EOF velocity with voxel size of 0.02 $\mu$m and a 0.02 V of electric potential. The Debye length is influenced by ionic concentration. When the ionic concentration is high enough, the Debye length becomes much smaller than the voxel size. The EDL is unresolvable and, therefore, the velocity profile appears plug-like. When the ionic concentration is small, the Debye length is comparable to the voxel size. Therefore, the EDL is resolvable and the velocity profile changed from plug-like to parabolic-like.}
    \label{fig:EOF_Cb}
\end{figure}

For the EOF validation in a 2D micro channel, as shown on the left of  Fig. \ref{fig:Validation}, two parallel plates are located at X = $\pm$ 0.5H, the domain is uniformly filled with a 1:1 electrolyte solution (NaCl). For the input micro channel to the LBPM and COMSOL, the pixel element is set to be 0.001 $\mu$m. Therefore, the total L and H of the channel are 0.1 $\mu$m and 0.05 $\mu$m, respectively. A meshed domain as the input for COMSOL is shown on the right of Fig. \ref{fig:Validation}. Debye length plays an essential role in the flow behavior for electroosmosis, which is a function of ionic concentration. Fig. \ref{fig:EOF_Cb} demonstrates how the ionic concentration affects the Debye length and therefore influences the velocity profile. For the region where the EDL is resolvable, the velocity profile is parabolic-like. With an increase of ionic concentration, the EDL becomes unresolvable, and the velocity profile changes from parabolic-like to plug-like.

In the following validation, the EK model in LBPM is validated against COMSOL in both scenarios where the EDL is resovable and unresolvable. To meet the condition where the EDL is fully resolvable (M = $\frac {H} {\lambda_{EDL}}$>0.01) \cite{zhang_electro-osmosis_2017}, the bulk ionic concentration is set to be 0.1$mol/m^3$. The Debye length is determined as 0.0302$\mu$m and M = 1.656 (in the range of fully resolvable). The surfaces of the plates are charged with a constant charge density, $\sigma$. For EOF, an electric potential (E) is imposed along the channel in the z-direction. Coulomb’s force (electromigration) acts on the electrolyte solution through the net charge within the EDL, and consequently the EOF occurs in the z-direction. Several EOFs under different $\sigma$ and E at steady state are compared between LBPM and COMSOL. 

\begin{figure}[htp!]
  \centering
    \includegraphics[width=\textwidth]{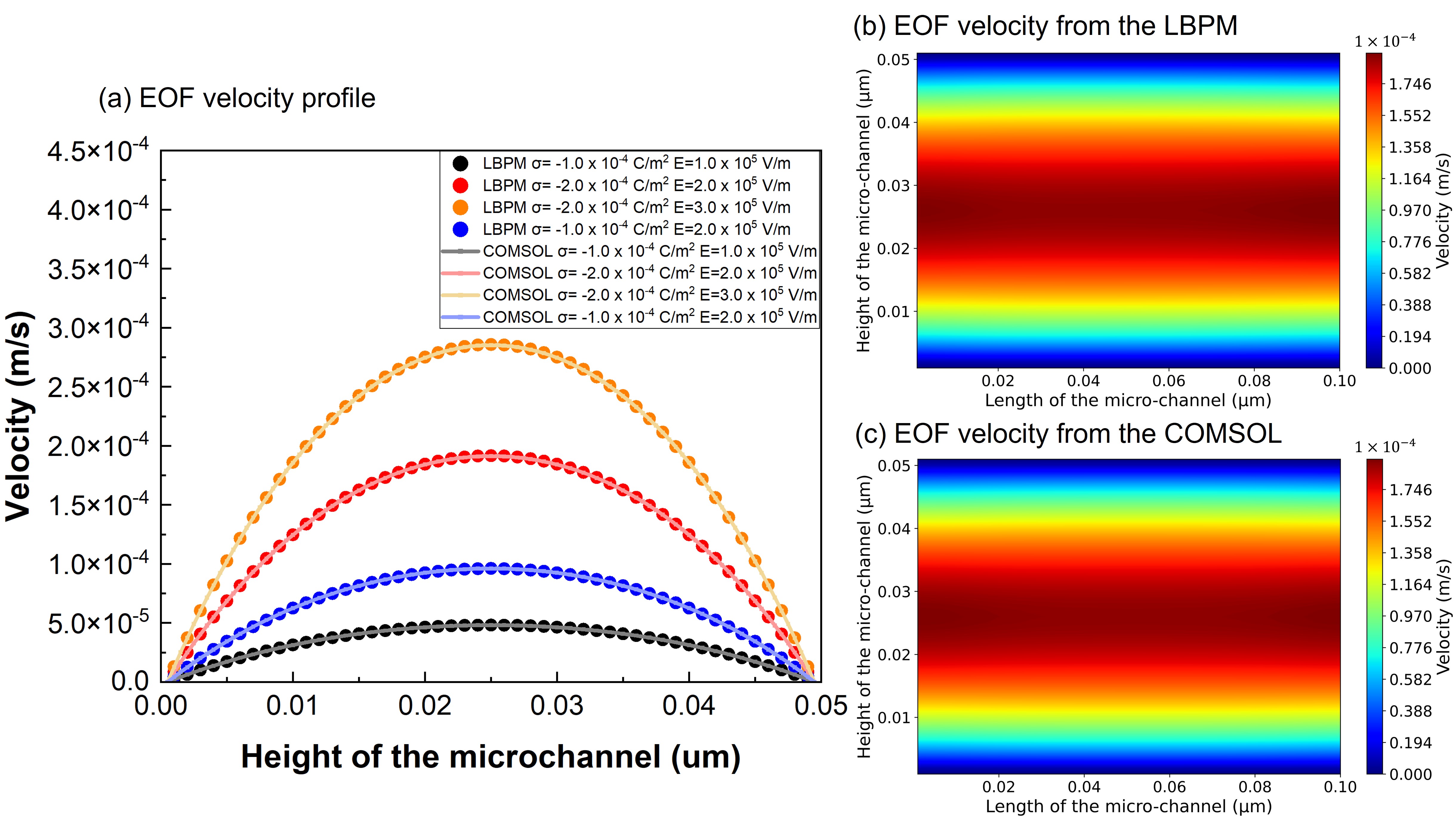}
    \caption{(a) Fully resolvable EOF validation results for the LBPM against the COMSOL under varied conditions. Good agreements are achieved in all cases. (b-c) The EOF velocity profile under the condition that EDL is fully resolvable for $\sigma = -2e-4C/m^2, E = 1e+5V/m$. (b) The velocity profile generated from LBPM. (c) The velocity profile generated from COMSOL. Good agreement is found in the velocity at all locations of the domain.}
    \label{fig:EOF_validation}
\end{figure}

Fig. \ref{fig:EOF_validation} (a) shows the parabolic-like velocity profile that is the average velocity along the z-direction. The velocity profile of the LBPM and COMSOL show good agreement in all cases where $\sigma$ = $-1\times10^{-4}C/m^2$, $-2\times10^{-4}C/m^2$ and E = $1\times10^{5} V/m$, $2\times10^{5} V/m$, $3\times10^{5} V/m$. The EOF is more significant under a large electric potential and surface charge, and thus results in a larger fluid velocity. The parabolic-like velocity profile matches the typical velocity profile when the EDL is fully resolvable (M>1) \cite{zhang_electro-osmosis_2017}. A visualization of the whole velocity domain is provided in Fig. \ref{fig:EOF_validation} (b-c) to further demonstrate that the LBPM result achieves good agreement against the COMSOL result for the entire velocity domain.    

\begin{figure}[H]
  \centering
    \includegraphics[width=\textwidth]{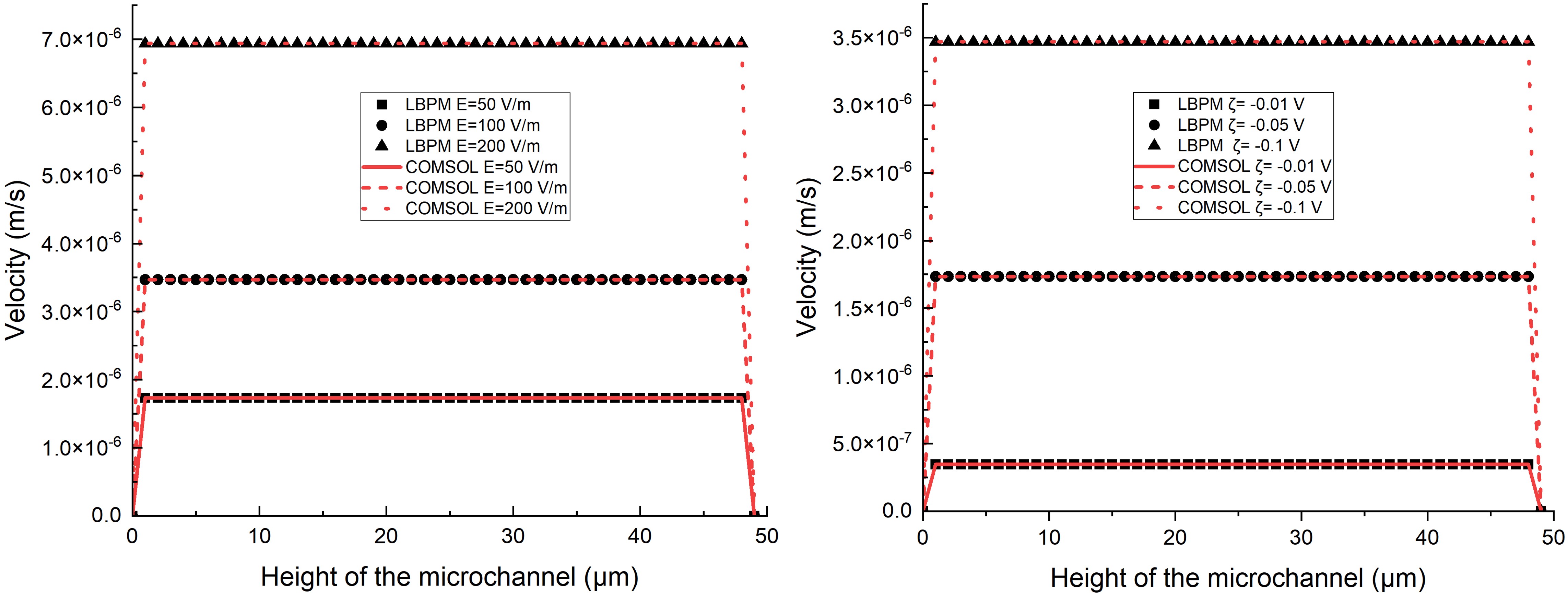}
    \caption{Left: EOF validation results when the EDL is unresolvable for LBPM against COMSOL, under varied electric potential gradient. Right: EOF validation results when the EDL is unresolvable for LBPM against COMSOL under varied $\zeta$. Good agreements are achieved in all cases.}
    \label{fig:slipvelocity}
\end{figure}

After validating the fully resolved EOF, the following validation is when the EDL is unresolvable. In this case, the thickness of the EDL is much smaller than the microchannel width so that the charge effect within the EDL is negligible and the domain remains electroneutral. The Helmholtz-Smoluchowski (HS) slip velocity is used to describe the slipping condition of the wall as a function of external electric potential and zeta potential. To meet the requirement that the EDL thickness is much smaller than H, the pixel size is set to be 1$\mu$m (H = 50$\mu$m), which is within a typical range of micro-CT image resolution. The ionic concentration is 0.1$mol/m^3$, M is then calculated, which is 0.0006 ($\ll$ 0.01). The validation result is shown in Fig. \ref{fig:slipvelocity}. A piston-like velocity profile is obtained with the HS boundary condition. The same shock front velocity (electroosmosis slip velocity) is obtained from LBPM and COMSOL under varied zeta potential and electric potential.

\subsection{EK versus hydraulic pressure driven transport}
\label{sec:hydraulic_pressure}

\begin{figure}[htp!]
  \centering
    \includegraphics[width=\textwidth]{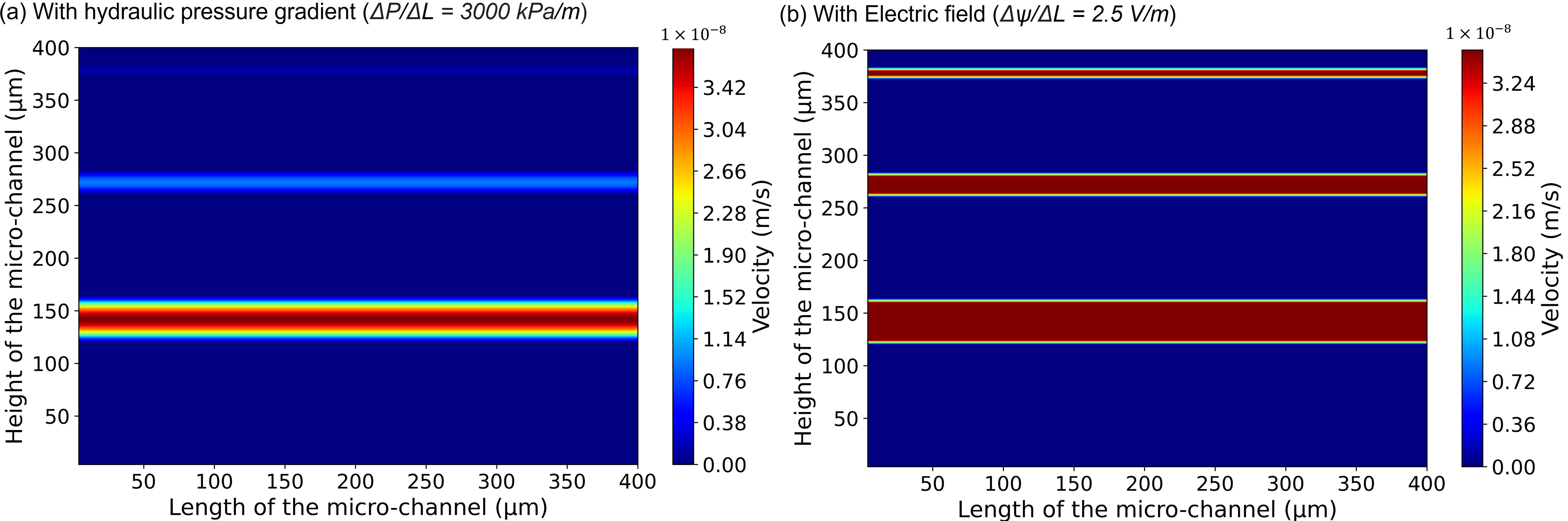}
    \caption{Single-phase flow of lixiviant in a heterogeneous micro-channel system under only (a) hydraulic pressure gradient or (b) electric field. Velocity profile shows that heterogeneous flow is obtained with hydraulic pressure, whereas a homogeneous flow is obtained with electric potential.}
    \label{fig:VelocityComp_pressure_EK}
\end{figure}

To understand the difference between single-phase flow behavior under either solely EK or hydraulic pressure, a single-phase flow simulation was performed using the validated EK model in a $100 \times 100$ 2D multi-microchannel system with a resolution of 4$\mu$m. The system contains only one type of solid phase with $\zeta = -0.02v$. Three microchannels with sizes of 8, 20, and 40$\mu$m were established in the system. The hydraulic pressure $\Delta$P was 3000$kPa/m$, and the electric field $\psi$ was 2.5$V/m$. The simulation settings of the lixiviant and diffusion coefficient are listed in Table \ref{tab:parameter} and \ref{tab:oreparameter}. For pressure-driven simulation, the external electric potential is set to be zero and only a pressure difference is applied.
The velocity profile results are shown in Fig. \ref{fig:VelocityComp_pressure_EK}. With only $\Delta$P, the flow was heterogeneous in the three microchannels, because with large channel size, the capillary entry pressure for fluid flow is lower than that of smaller channel size \cite{lake1989enhanced}. Therefore, fluid prefers to enter the largest-sized channel owing to less hydraulic resistivity, as shown in Fig. \ref{fig:VelocityComp_pressure_EK}(a). Consequently, the flow velocities were greater in the larger channel. For ISR, only minerals that resided in the larger pores and fractures with preferential flow would be recovered efficiently, whereas minerals in the smaller pores would be transported less efficiently. On the other hand, a promising flow behavior was obtained when applying an electric potential, as shown in Fig. \ref{fig:VelocityComp_pressure_EK}(b). The velocity in all three channels is homogeneous, and the velocity profile is sharp, indicating that all minerals in the pores or fractures can be efficiently exposed to lixiviant, independent of the pore size. Based on these results, it can be seen that flow driven by EK mechanism enables the minimization of the effect of pore-scale heterogeneity, thus promising a higher mineral recovery efficacy during ISR.      

\renewcommand{\arraystretch}{1.8}
\begin{table}[htp!]
\centering
\caption{Settings for numerical simulation, which are the same as the previously reported experimental conditions (M represents molarity). At inlet and outlet, the numerical settings are the boundary conditions, and in the system, the numerical settings are the initial condition.}
\begin{tabular}{lll}
\hline
\textbf{Solution} & \textbf{Experimental settings} & \textbf{Numerical settings}              \\ \hline
At the inlet             & 0.5$M$ $FeCl_{3}$ + 0.2$M$ $HCl$          & 0.5$M$ $Fe^{3+}$, 0.2$M$ $H^+$, 1.7$M$ $Cl^-$, 0$M$ $OH^-$, 0$M$ $Na^+$    \\
In the system     & 0.1$M$ $NaCl$                      & 0$M$ $Fe^{3+}$,$1\times10^{-7} M$ $H^+$, 0.1$M$ $Cl^-$, $1\times10^{-7} M$ $OH^-$, 0.1$M$ $Na^+$ \\
At the outlet            & 0.3$M$ $NaCl$ + 0.2M $HCl$           & 0$M$ $Fe^{3+}$, 0.2$M$ $H^+$, 0.5$M$ $Cl^-$, 0$M$ $OH^-$, 0.3$M$ $Na^+$ \\ \hline  
\end{tabular}
\label{tab:oreparameter}
\end{table}

\subsection{EK in a Chalcopyrite-Silica System}
\label{sec:realOre}

To apply EK-ISR, the migration of Fe$^{3+}$ through the ore system must be ensured for the dissolution of chalcopyrite into the fluid phase. Therefore, the discussion of Fe$^{3+}$ ion is focused on this section. Based on the validation results, the EK model in LBPM showed good agreement with COMSOL and was thus used for studying the EK transport in a complex chalcopyrite-silica system. To reduce the computational time for the simulation, a representative 256 cubic voxels subdomain was cropped from the original domain, as shown in Fig. S2 in the Supplemental Materials, Section 2. Representative elementary volume analysis of the chalcopyrite-silica system was also performed, and the results are provided in the Supplemental Materials. The typical chalcopyrite content in the sulfide ore is around 0.3-5 \% \cite{agar1991flotation,ikiz2006dissolution,velasquez2018leaching}. The mineral composition of the representative sub-domain that is used in the following simulation is 96.3\% silica and 3.7\% chalcopyrite, which is in the typical range. Moreover, to check the physical accuracy of the proposed EK model, and also considering the heterogeneous distribution of chalcopyrite in the small sub-section of the whole sample, two manually generated subdomains which have chalcopyrite concentration of 15.6\% and 63.3 \% were created and used to check the physical accuracy of the model and study the effect of zeta potential for different percentages of chalcopyrite. These subdomains were obtained by changing the silica grains to chalcopyrite grains. The material compositions for all simulated representative sub-domains are listed in Table \ref{tab:subdomains}. The following numerical studies were conducted on these sub-domains.

\renewcommand{\arraystretch}{1.5}
\begin{table}[htp!]
\centering
\caption{Mineral compositions of all three sub-domains used in this study. Domain1 is the real scanned sub-domain. Domain2 and Domain3 are the manually generated sub-domains.}
\begin{tabular}{llll}
\hline
\textbf{}   & Domain1   & Domain2   & Domain3   \\ \hline
Voxel size  & 256 cubic & 256 cubic & 256 cubic \\ \hline
Chacopyrite & 3.7\%     & 15.6\%    & 63.3\%    \\ \hline
Silica      & 96.3\%    & 84.4\%    & 36.7\%    \\ \hline
\end{tabular}
\label{tab:subdomains}
\end{table}

\subsubsection{EK under Different Zeta Potentials}
\label{sec:zetapotential}

The simulation settings were set to match the EK-ISR laboratory-scale experimental settings in \cite{martens_toward_2021}, as shown in Table \ref{tab:oreparameter}. At inlet and outlet, the 0.2M HCl is set as boundary condition. In the system, $1\times10^{-7} M$ is set as initial condition. During the simulation, $H^{+}$ is transported into the system under diffusion and electromigration. Chalcopyrite dissolution by the lixiviant ($FeCl_{3}$), which is the reaction with dissolved $Fe^{3+}$ is described as follows:

\begin{equation}
    \ce{CuFeS_{2} + 4Fe^{3+} -> Cu^{2+} + 5Fe^{2+} + 2S^{0}}.
\end{equation}

The oxidative dissolution of chalcopyrite occurs due to the reaction with $Fe^{3+}$ ions present in the lixiviant. Under an electric field, $Fe^{3+}$ moves from the source reservoir to the copper bearing reservoir, where $Cu^{2+}$ is leached. The dissolved $Cu^{2+}$ and other cations are then transported to the target reservoir via electromigration. The reactive surface area plays a fundamental role on the mineral dissolution and reactive transport processes. The porous media properties such as tortuosity, porosity and permeability evolve as the mineral dissolution takes place. Such changes in the pore structure alter the magnitude of the velocity field and control the flow pattern. Moreover, mineral dissolution modifies the local pH in the system, affecting the zeta potential which has a great influence on the electroosmotic permeability. Therefore, understanding $Fe^{3+}$ transport in the domain is non-trivial. The pH at the inlet and outlet was set to 0.7 (initial strongly acidic conditions). Other settings are listed in Table S1. The zeta potential $\zeta$ for silica and chalcopyrite is another important factor that controls the fluid/ion flow behavior. Moreover, transport is sensitive to the pH of the solution. For silica and chalcopyrite, experimental measurements of $\zeta$ at various pH values have been reported in \cite{xu_preparation_2006,runqing_surface_2010}. Under pH around 0.7, the $\zeta$ value for silica was approximately $-0.005V$. For chalcopyrite, $\zeta$ was approximately $0.002V$, whereas $\zeta$ of chalcopyrite after the treatment with ferric chromium lignin sulfonate is approximately $-0.03V$. To further investigate the effect of $\zeta$ on EK, three simulations using three subdomains (3.7, 15.6, and 63.3\% chalcopyrite) with $\zeta = 0.002V$ were compared.

\begin{figure}[htp!]
  \centering
    \includegraphics[width=0.95\textwidth]{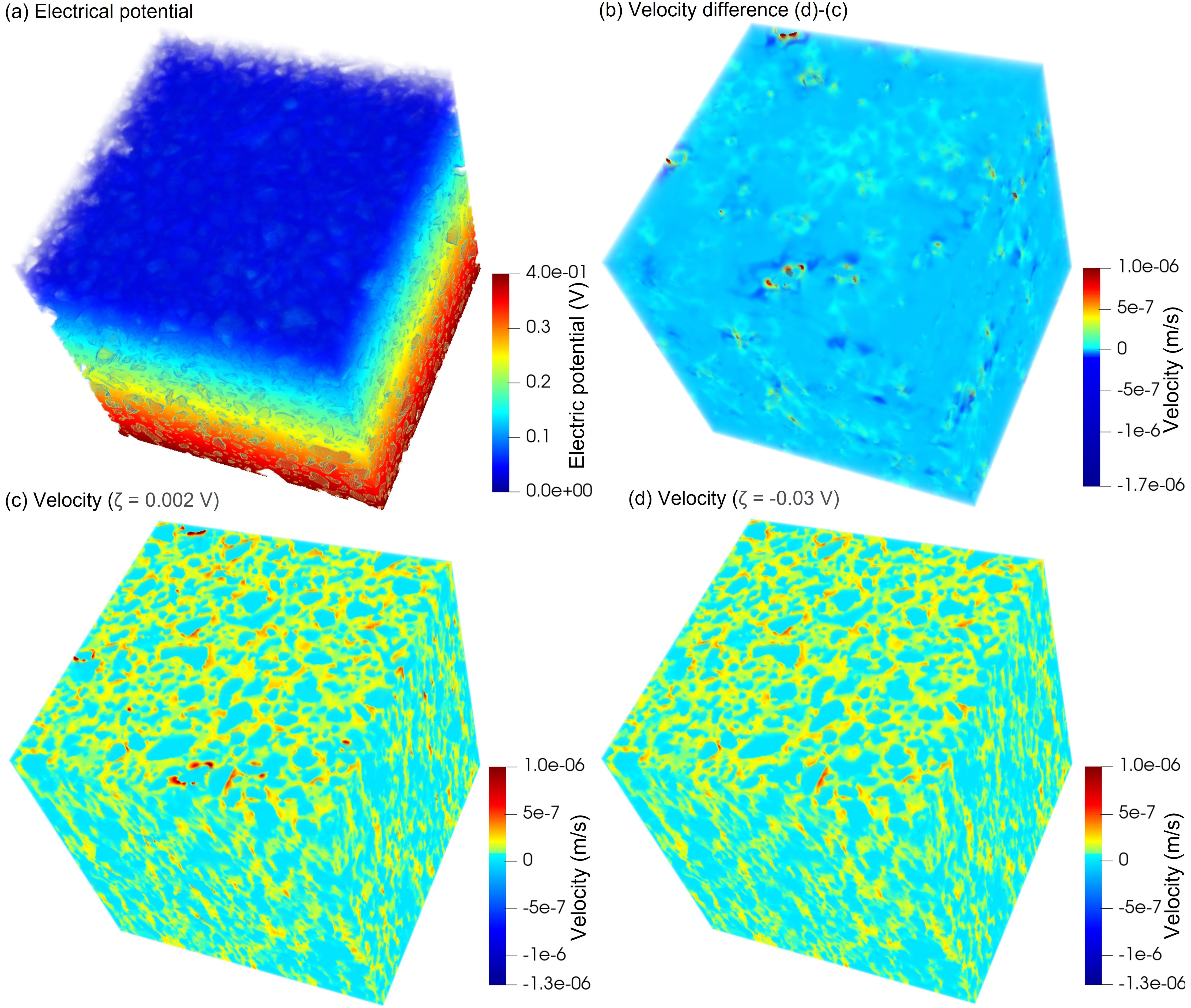}
    \caption{Simulation results for Domain1: (a) electric potential; (b) fluid velocity difference between chalcopyrite $\zeta$ of $0.002 V$ and $-0.03 V$; (c) fluid velocity profile of $\zeta = 0.002 V$; (d)  fluid velocity profile of $\zeta = -0.03 V$.}
    \label{fig:velocity_psi}
\end{figure}

With a high ionic concentration and large voxel size (5.4 $\mu$m), the EDL fell into the unresolvable regime; therefore, Eq.\ref{eq:HSequation} was used to compute the flow velocity. Using the HS equation, the velocity rapidly converged. The stopping criterion for the steady-state condition was the average velocity difference between the two time steps less than $10^{-9} m/s$. Figure \ref{fig:velocity_psi} shows the electric potential, velocity profiles for $\zeta = 0.002V$ and $\zeta = -0.03V$, and the difference between the resulting velocity profiles. Electroosmosis is related to the $\zeta$ value of each mineral. For a negative $\zeta$, the fluid moves towards the cathode, whereas for a positive $\zeta$, the fluid moves towards the anode. The major mineral in the subdomain was silica with $\zeta = -0.005 V$. The fluid surrounding the silica flowed towards the cathode (positive velocity value). For $\zeta$ of chalcopyrite of $0.002 V$, the fluid flowed towards the anode (negative velocity value), indicating that electroosmosis acted in the opposite direction to electromigration. However, when $\zeta$ for chalcopyrite was $-0.03V$, the fluid surrounding the chalcopyrite flowed at a higher velocity than that surrounding the silica, indicating that the effects of electroosmosis and electromigration were codirectional. The average velocities and electroosmotic permeabilities are listed in Table \ref{tab:EOFperm_zeta}. The average velocity for $\zeta = -0.03V$ was slightly higher than that for $\zeta = 0.002V$, owing to the EOF. The electroosmotic permeability governs the fluid flow under the electric potential similarly to how the hydraulic conductivity governs the flow under a hydraulic gradient. Therefore, a larger electroosmotic permeability was obtained for a more negative $\zeta$. The difference is 7.28\% because the amount of chalcopyrite in Domain1 is small compared to that in silica. 

To investigate the effect of chalcopyrite occurrence, Domain2 and Domain3 were compared with Domain1 under the same conditions, where $\zeta = 0.002 V$. The results are presented in Table \ref{tab:EOFdif_fraction}. Ion flux owing to advection, diffusion, and electromigration are also reported. With an increase in chalcopyrite fraction, the average velocity decreases significantly (up to 76.21\% difference between Domain1 and Domain3) because the velocity near the chalcopyrite is negative owing to the positive zeta potential, leading to a decrease in the average velocity. For $Fe^{3+}$ ion flux, with an increase in chalcopyrite fraction, the advective flux decreases because advection is related to fluid velocity. However, the diffusive flux remained almost constant for all the three domains, whereas the electrical flux slightly decreased. Comparing the ionic flux, the flux of advection due to EOF is 4 magnitudes less that electromigration and diffusion. Péclet number ($Pe$) that calculates the ratio between advection transport and diffusion transport is also reported for all ions, as defined in Eq. (1) and (2) in Supplementary materials. $Pe$ provides an indication of the dominant transport mechanism in the system independent of the geometry size of the domain. The advection term in $Pe$ is contributed by EOF and electromigration. The $Pe$ result is listed in Table 1 in Supplementary materials. The $Pe$ shows that for all ions (except Fe$^{3+}$), the $Pe$ is around 1, while for Fe$^{3+}$, the $Pe$ is around 3. The EOF and eletromigration induced velocity value is also reported. Overall, the results reveal that for EK-ISR, where an external electric potential is applied, electromigration is the dominant ion transport mechanism. The diffusive flux is the second important transport mechanism in our simulation settings because the domain initially contained no $Fe^{3+}$, resulting in high diffusive flux. EOF contributes few to the ions transport.

\renewcommand{\arraystretch}{1}
\begin{table}[H]
\centering
\caption{Electroosmotic permeability results at $\zeta = 0.002 V$ and $-0.03 V$ for Domain1. The average velocity at $\zeta = -0.03 V$ is greater than at $\zeta = 0.002 V$. Because the same external electric potential is applied, the electroosmotic permeability at $\zeta = -0.03 V$ is greater than at $\zeta = 0.002 V$.}
\begin{tabular}{lll}
\hline
\textbf{}    & $\zeta = -0.03 V$ & $\zeta = 0.002 V$ \\ \hline
Average Velocity ($m/s$)                & $1.51\times10^{-7}$           & $1.40\times10^{-7}$  \\
electric potential gradient ($V/m$)       & $2.89\times10^{2}$           & $2.89\times10^{2}$  \\
Electroosmotic permeability $(m^{2}/(V \cdot s))$ & $5.23\times10^{-10}$   & $4.84\times10^{-10}$ \\ \hline
\end{tabular}
\label{tab:EOFperm_zeta}
\end{table}

\renewcommand{\arraystretch}{1}
\begin{table}[H]
\centering
\caption{Average velocity and ferric ion flux under advection, diffusion, and electromigration for Domain1, Domain2, and Domain3 with $\zeta=0.002V$ for chalcopyrite and $\zeta=-0.005V$ for silica.}
\begin{tabular}{llll}
\hline
Chalcopyrite fraction        & 3.7\%        & 15.6\%     & 63.3\%     \\ \hline
Average   velocity ($m/s$)    & $1.40 \times 10^{-7}$     & $1.14 \times 10^{-7}$   & $3.33\times10^{-8}$   \\
Percentage of difference    &              & 18.57\%    & 76.21\%    \\ \hline
Ion advective flux ($mol/s$)  & $1.2677\times10^{-18}$  & $1.0318\times10^{-18}$ & $3.1145 \times 10^{-19}$ \\
Percentage of difference    &              & 18.61\%    & 75.43\%    \\ \hline
Ion diffusive flux ($mol/s$)  & $4.8877\times10^{-14}$  & $4.8873\times10^{-14}$ & $4.8873\times10^{-14}$ \\
Percentage of difference    &              & 0.01\%     & 0.01\%     \\ \hline
Ion electrical flux ($mol/s$) & $9.0835\times10^{-14}$ & $9.0833\times10^{-14}$ & $9.0827\times10^{-14}$ \\
Percentage of difference    &              & 0.002\%    & 0.01\%    \\\hline
\end{tabular}
\label{tab:EOFdif_fraction}
\end{table}

\begin{figure}[htp!]
  \centering
    \includegraphics[width=\textwidth]{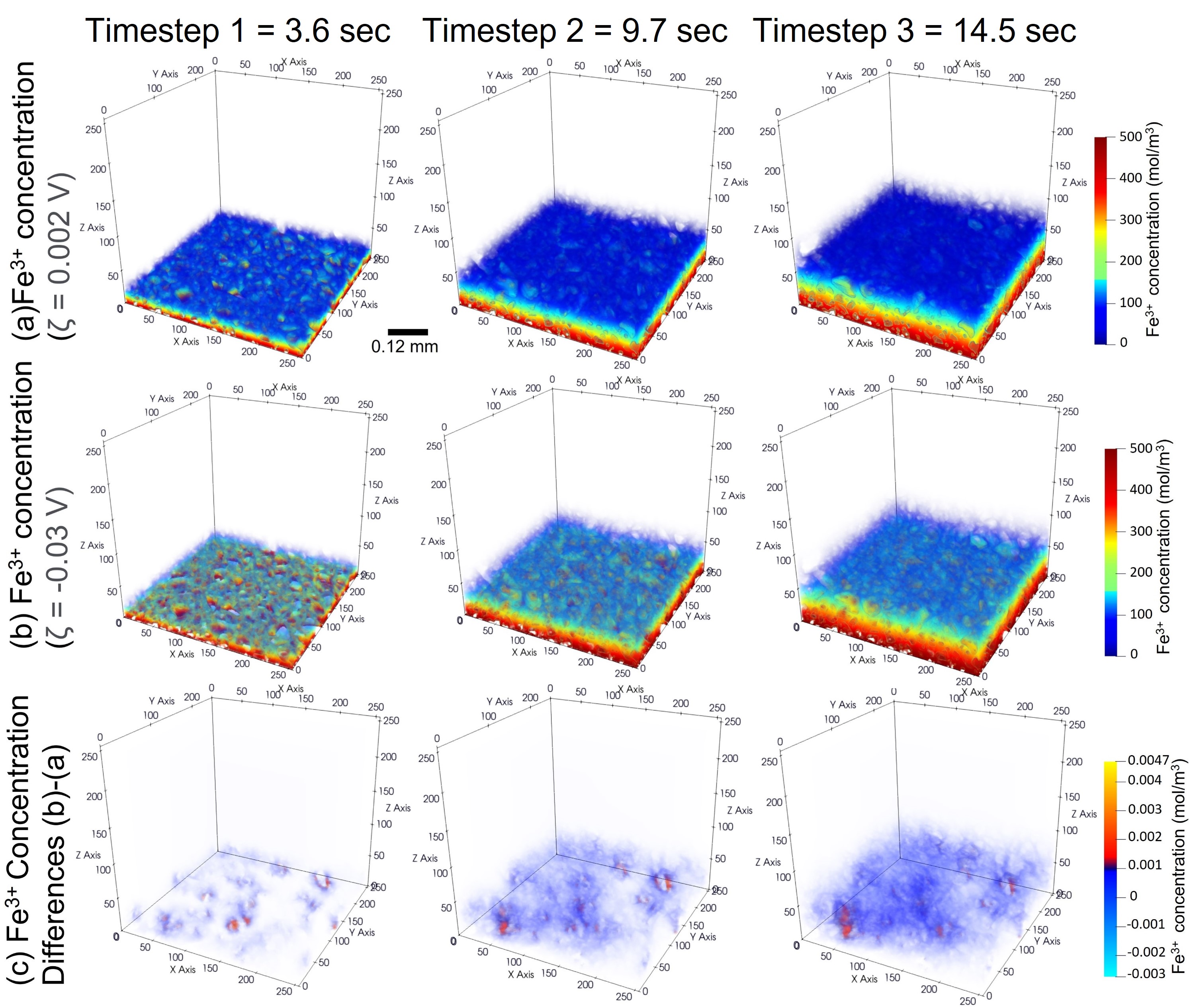}
    \caption{Fe$^{3+}$ distribution in Domain1 under different $\zeta$ of chalcopyrite: (a) $\zeta = 0.002 V$; (b) $\zeta = -0.03 V$. (c) Differences in Fe$^{3+}$ distribution between (a) and (b).}
    \label{fig:Feconcentration}
\end{figure}

Under an external electric potential, positively charged ions move towards the cathode, and negatively charged ions move towards the anode owing to electromigration. Fe$^{3+}$, therefore, moves from the anode to the cathode, as shown in Timesteps 1--3 in Fig. \ref{fig:Feconcentration} (a) and (b). The physical time corresponding to them is 3.6, 9.6, and 14.5sec, respectively. which illustrates the transport of Fe$^{3+}$ ions under applied electric potential, . As Fe$^{3+}$ moves into the ore system, a larger region of chalcopyrite is exposed, which is essential for the reaction and recovery of Cu$^{2+}$. Fig. \ref{fig:Feconcentration} (c) calculates the differences in Fe$^{3+}$ distribution between Fig. \ref{fig:Feconcentration} (a) and Fig. \ref{fig:Feconcentration} (b), and shows the different Fe$^{3+}$ concentration due to different zeta potential at mineral surface. Different zeta potential results in different velocity profile calculated from Eq. \ref{eq:NavierStokesEqs} and Eq. \ref{eq:HSequation}. This velocity causes the difference in Fig. \ref{fig:Feconcentration} (c) by the advection flow shown in Eq. \ref{eq:NernstPlanckEq}. Overall, pH and the geochemical reactions involved with $\zeta$ for different minerals in the ore are essential parameters affecting the EK performance, and therefore needs to be comprehensively characterized.

\subsubsection{EK Under Different Electric Potentials}
\label{sec:electricalpotential}
\begin{figure}[htp!]
  \centering
    \includegraphics[width=0.9\textwidth]{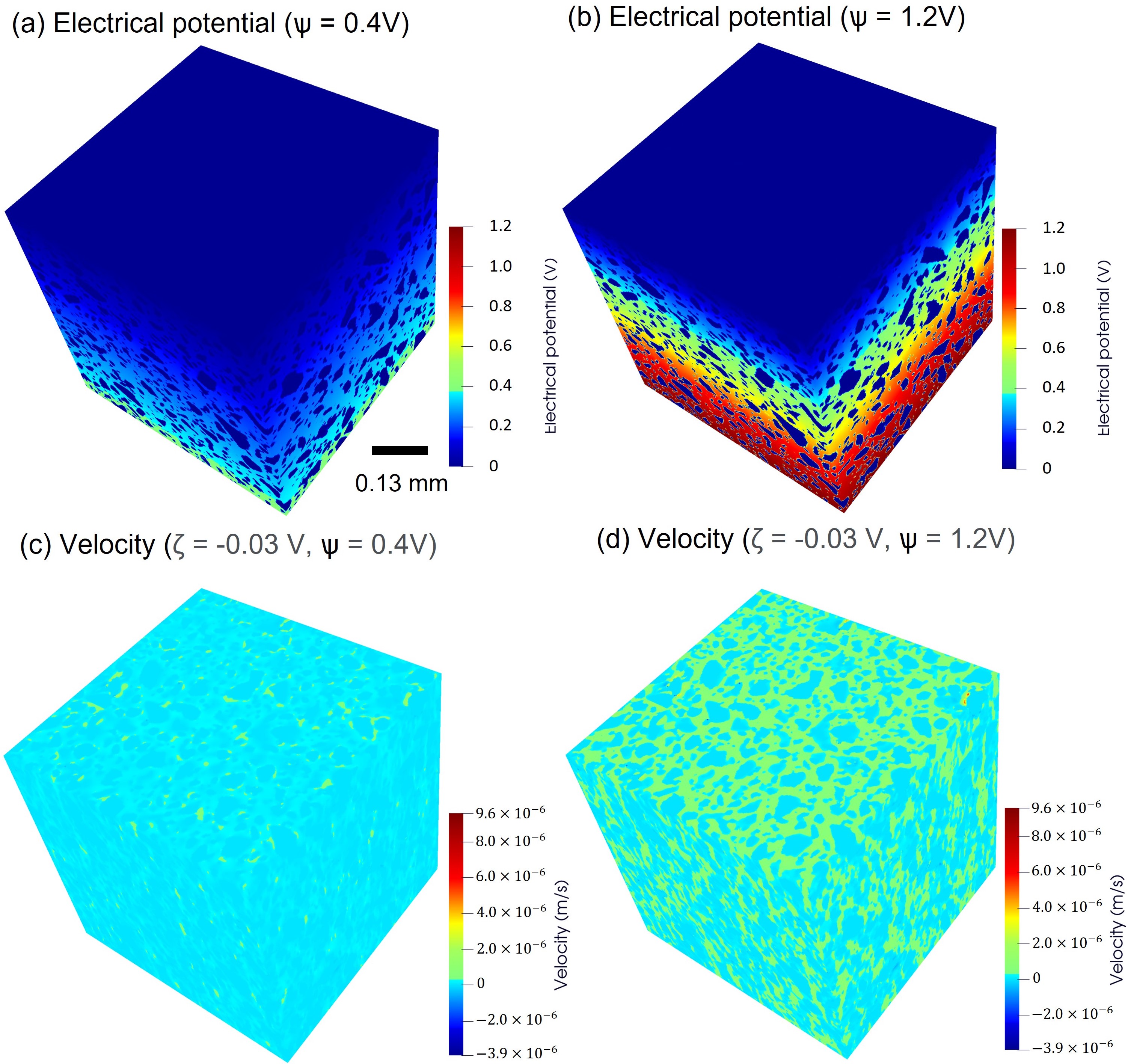}
    \caption{Simulation results for Domain1: (a) electric potential at $\psi = 0.4 V$; (b) electric potential at $\psi = 1.2 V$. (c) Velocity profile of $\psi = 0.4 V$. (d) Velocity profile of $\psi = 1.2 V$.}
    \label{fig:velocity_1.2V_0.4V}
\end{figure}

Based on previous results, electromigration is the main transport mechanism in EK-ISR. Therefore, simulations with different electric potentials ($\psi = 0.4 V$ and $\psi = 1.2 V$) were computed at $\zeta = -0.03 V$ for chalcopyrite and $\zeta = -0.005 V$ for silica. Fig. \ref{fig:velocity_1.2V_0.4V} (a) and (b) show the electric potential for both cases. Figs. \ref{fig:velocity_1.2V_0.4V} (c) and (d) show the z-component velocity for the two cases. The velocity towards the cathode under a higher electric potential was greater than that under a lower electric potential at negative $\zeta$ for both silica and chalcopyrite. This can be explained using Eq. \ref{eq:HSequation}, where the velocity is proportional to the electric potential. The average velocity exhibited the same trend, as listed in Table \ref{tab:EOFperm_psi}. Notably, if $\zeta$ is positive for chalcopyrite, under a higher electric potential, the average velocity towards the cathode becomes smaller than that under a lower electric potential. This is because at positive $\zeta$, electroosmosis leads to flow in the opposite direction (negative velocity) towards the anode at the locations surrounding chalcopyrite. Additionally, the electroosmotic permeability is the same in both cases, which indicates that the electroosmotic permeability is independent of the external electric potential. This condition is valid for this case because a constant of 0.2$M$ HCl was applied at the inlet and outlet, resulting in a homogeneous pH distribution in the system. However, this condition is not met when there is a variation in pH. For example, \cite{zhang_electro-osmosis_2017} found that electroosmotic permeability can change under heterogeneous pH conditions. 


\renewcommand{\arraystretch}{1}
\begin{table}[H]
\centering
\caption{Electroosmotic permeability results at $\psi = 1.2 V$ and $0.4 V$ for Domain1. The average velocity at $\psi=1.2V$ is greater than at $\psi = 0.4 V$. The electroosmotic permeability, however, is same for both $\psi$.}
\begin{tabular}{lll}
\hline
\textbf{}    & $\psi = 0.4 V$ & $\psi = 1.2 V$ \\ \hline
Average Velocity ($m/s$)                & $1.51\times10^{-7}$         & $4.54\times10^{-7}$  \\
electric potential gradient ($V/m$)       & $2.89\times10^{2}$           & $8.68\times10^{2}$  \\
Electroosmotic permeability $(m^{2}/(V \cdot s))$ & $5.23\times10^{-10}$   & $5.23\times10^{-10}$ \\ \hline
\end{tabular}
\label{tab:EOFperm_psi}
\end{table}

\begin{figure}[htp!]
  \centering
    \includegraphics[width=\textwidth]{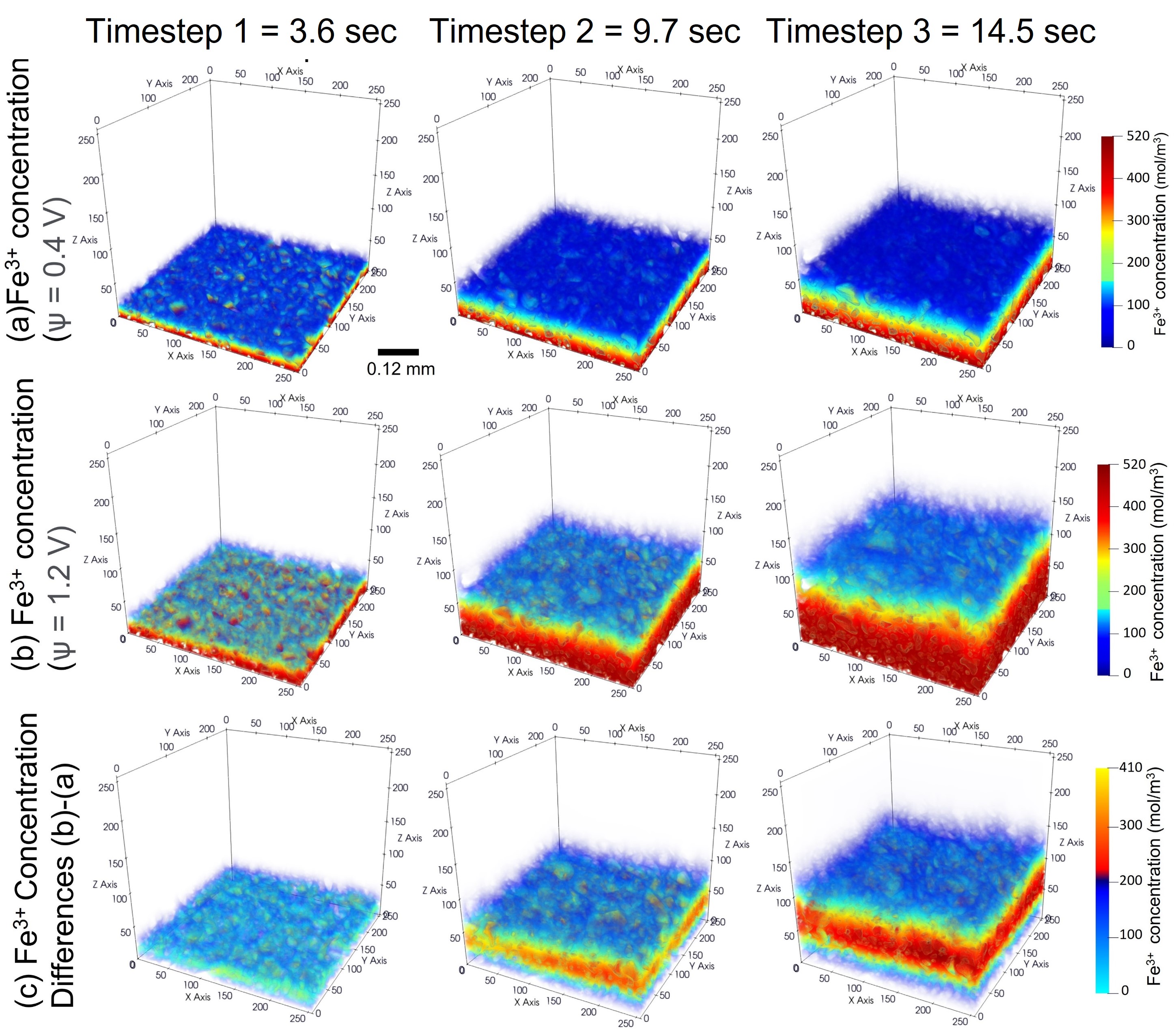}
    \caption{$Fe^{3+}$ distribution in Domain1 under different external $\psi$: (a) $\psi = 0.4 V$; (b) $\psi = 1.2 V$. (c) Differences in $Fe^{3+}$ distributions between (a) and (b).}
    \label{fig:FeconcentrationEdifferent}
\end{figure}

Although the electric potential has no effect on electroosmotic permeability, it dominates the distribution of $Fe^{3+}$ in the system. Fig. \ref{fig:FeconcentrationEdifferent} (a) and (b) show the distributions of $Fe^{3+}$ concentration at three different timesteps under electric potentials of $\psi = 0.4 V$ and  $\psi = 1.2 V$, respectively. Fig. \ref{fig:FeconcentrationEdifferent} (c) shows the difference between the concentration fields. From timestep 1 to 3, $Fe^{3+}$ ions under $\psi = 1.2 V$ flowed faster from the anode to the cathode than those under $\psi = 0.4 V$. This resulted in a broader distribution of $Fe^{3+}$ ions in the sample, which is beneficial for chalcopyrite dissolution.  

\begin{figure}[htp!]
  \centering
    \includegraphics[width=\textwidth]{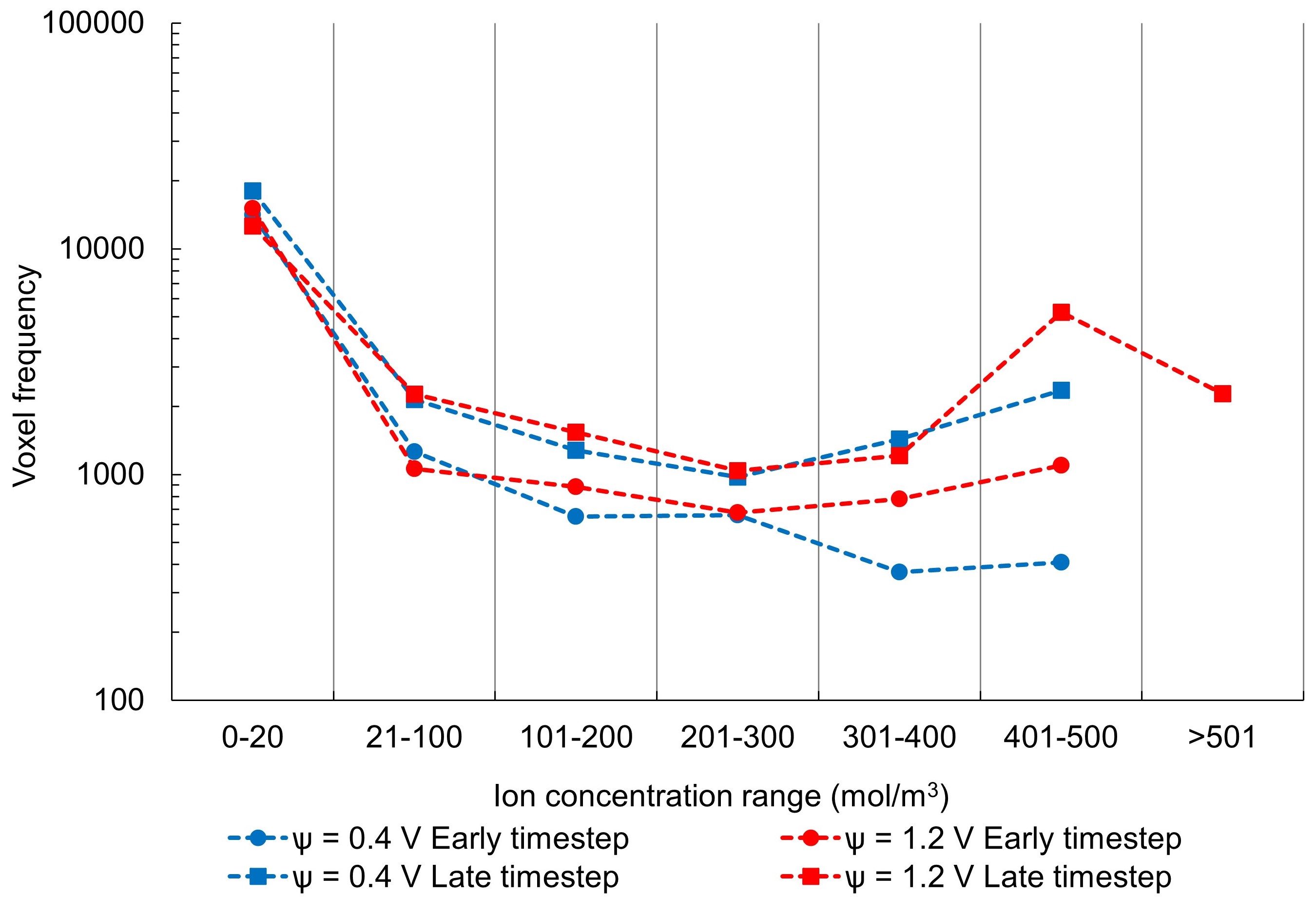}
    \caption{$Fe^{3+}$ distribution is divided into seven concentration ranges. The voxel frequency refers to the number of voxels that are connected to the voxels of chalcopyrite and also fall into the satisfied concentration range. The results in early and late timesteps are compared for $\psi = 0.4 V$ and $\psi = 1.2 V$. For the range $>501 mol/m^{3}$, except for the late timestep for $\psi = 1.2 V$, the voxel frequency for others are 1.}
    \label{fig:Fe2Cu}
\end{figure}

Fig. \ref{fig:Fe2Cu} shows the relationship between $Fe^{3+}$ and chalcopyrite at different timesteps. When $\psi = 0.4 V$, at an early timestep, the majority of $Fe^{3+}$ concentration near the chalcopyrite is in the range of 0 to 20 $mol/m^{3}$, and only one voxel has an $Fe^{3+}$ concentration of $>501 mol/m^{3}$. At the late timestep, the number of voxels in the range 0 to 20 $mol/m^{3}$ decreases, whereas the number of voxels in other ranges increases, which is especially significant for the $Fe^{3+}$ concentration of $>501 mol/m^{3}$ (from 1 to 2278). When $\psi$ increases to 1.2$V$, as the simulation progresses, more voxels have $Fe^{3+}$ concentration higher than 21 $mol/m^{3}$, particularly with an increase in $Fe^{3+}$ concentration greater than 401 $mol/m^{3}$. $Fe^{3+}$ moves faster at higher $\psi$. In both cases, as the timestep increases, more $Fe^{3+}$ flows into the system and contacts the chalcopyrite body.

\section{Conclusions}
\label{sec:conclusions}

In this study, we propose a pore-scale EK model built with LBPM to describe the advection/diffusion, electromigration and electroosmosis of fluid and charged species in the complex porous media. Key features of the proposed model are: (1) capable of simulating EK on complex porous media of images; (2) characterizse the fluid/ion flow under both condition when EDL is resolvable and unresolvable; (3) support GPU-acceleration and therefore, can be used for large domain simulation as a digital twin study with of experiments result; and (4) the model that solves the transport of fluid and charged species can be coupled to PhreeqcRM in Python and MATLAB. The EK model was validated against COMSOL Multiphysics in terms of EOF in a 2D microchannel. The thickness of the EDL and its effect on the numerical model were discussed and validated. Specifically, at the thickness of EDL comparable to the domain size, the EDL can be fully resolved by the EK model, whereas at the thickness of EDL smaller than the domain size, the EDL is unresolvable by the model. Therefore, the HS equation was used to compute the slip boundary velocity. Good agreement was obtained between the EK model and COMSOL Multiphysics.

Subsequently, a chalcopyrite-silica system consisting of chalcopyrite and silica powder was prepared and imaged with micro-CT. The simulation with the chalcopyrite-silica system provides a more complex porous structure for characetrising fluid and ions flow under the EK condition. The simulation application studied the EK processes under various mineral occurrence, zeta potential, and electric potential. The results highlight the important influence of mineral occurrence, zeta potential, and electric potential to electroosmosis and electromigration for EK. The flexibility of the model opens the opportunity to coupling the surface complexion model for local pH characterization, such as 1-pk model \cite{de2010three} and triple layer model \cite{revil2004constitutive}, coupling with a geochemical model, such as PhreeqcRM to fully capture the reactive transport under electric potential \cite{parkhurst2015phreeqcrm}, and simulating large domain ore sample which is used in experiments. Such model application can be applied for EK-ISR where the low permeability-porosity system is presented \cite{martens_toward_2021}.

\section{Funding}
The authors did not receive support from any organization for the submitted work. The authors have no relevant financial or non-financial interests to disclose.

\section{Acknowledgment}
\label{sec:Acknowledgement}
The authors acknowledge the Tyree X-ray CT Facility, UNSW network lab, funded by the UNSW Research Infrastructure Scheme.

\bibliographystyle{ieeetr}  
\bibliography{LBPM_COMSOL_paper}

\end{document}


\setcounter{tocdepth}{4}
\setcounter{secnumdepth}{4}
\maketitle

\pagebreak
\section{CNN Architecture and Training Schedule}
\label{sec:cnn}

\begin{figure}[H]
  \centering
    \includegraphics[width=\textwidth]{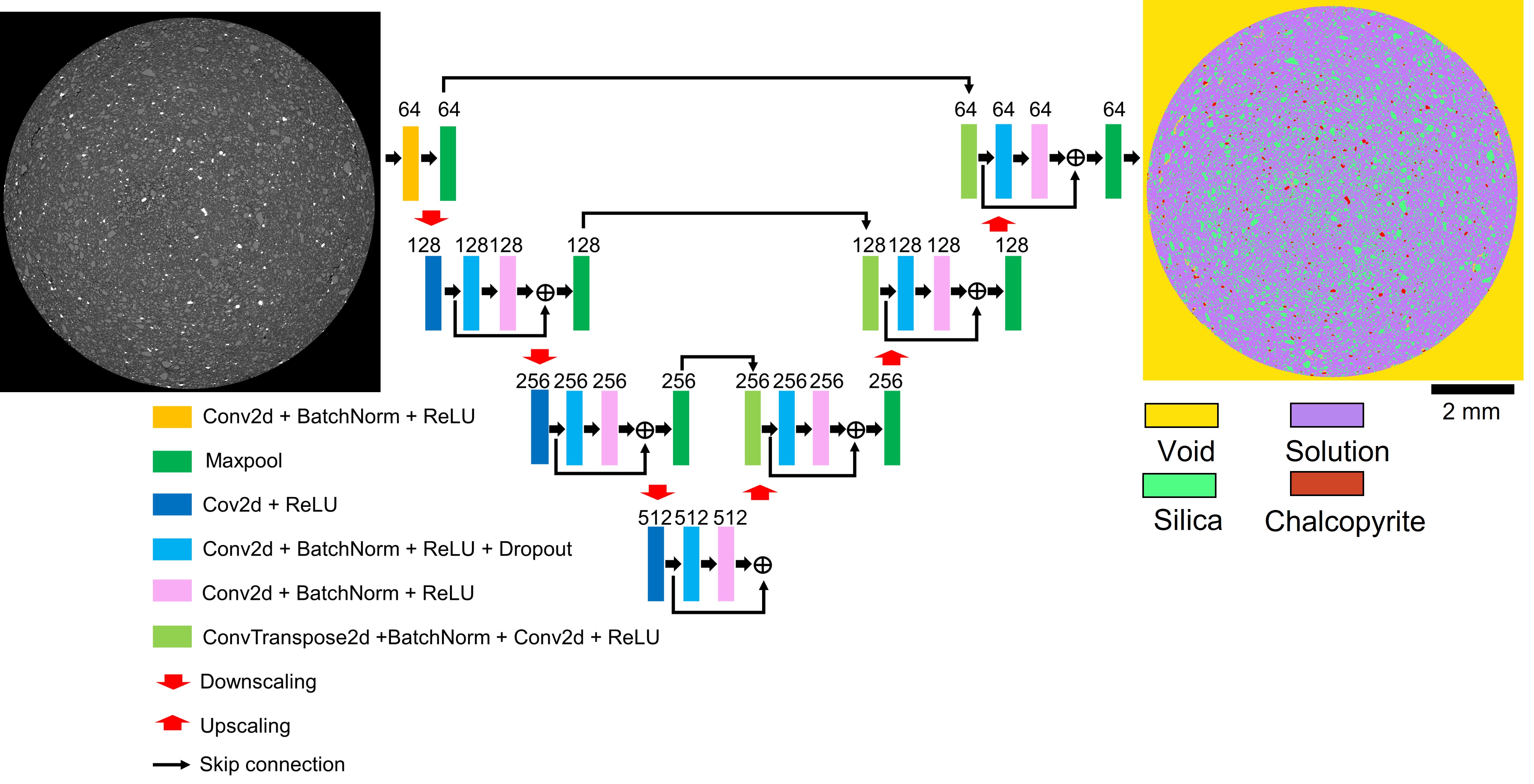}
    \caption{Architecture of the U-ResNet used for segmentation, containing downsampling process as image feature extraction and upsampling layer for useful feature decoding.}
    \label{fig:CNN}
\end{figure}

Training data for U-ResNet \cite{wang_deep_2021} (architecture was shown in Fig. \ref{fig:CNN} contained the grayscale micro-CT slice and its corresponding ground truth from the WEKA segmentation tool with a size of $2013 \times 2013$ pixels. It was cropped into patches with a size of $100 \times 100$ pixels. Overall, 400 patches were obtained, and split 80/20 for training and testing. The CNN model was trained for 100 epochs with an initial learning rate of 0.0001 and batch size of 16 using the Adam solver \cite{kingma_adam_2017}. The learning rate was reduced by a factor of 0.5 each time the loss reaches a plateau after ten epochs. The training was implemented in PyTorch, using an Nvidia RTX 3090 GPU and 16 cores' CPU (AMD 5950X) with a memory of 128 GB. The approximate training time per epoch was 5 seconds, the model was trained in 9 minutes for 100 epochs. After training, other 2969 full-size 2D slices were passed to the CNN model one-by-one for segmentation. For each full size slice segmentation, the approximate training time was 12 seconds. Therefore, 10 hours were used to segment the whole 3D synthetic ore system.
 
\section{Synthetic ore system and homogeneous analysis}
\label{sec:ore}
\begin{figure}[htp!]
  \centering
    \includegraphics[width=\textwidth]{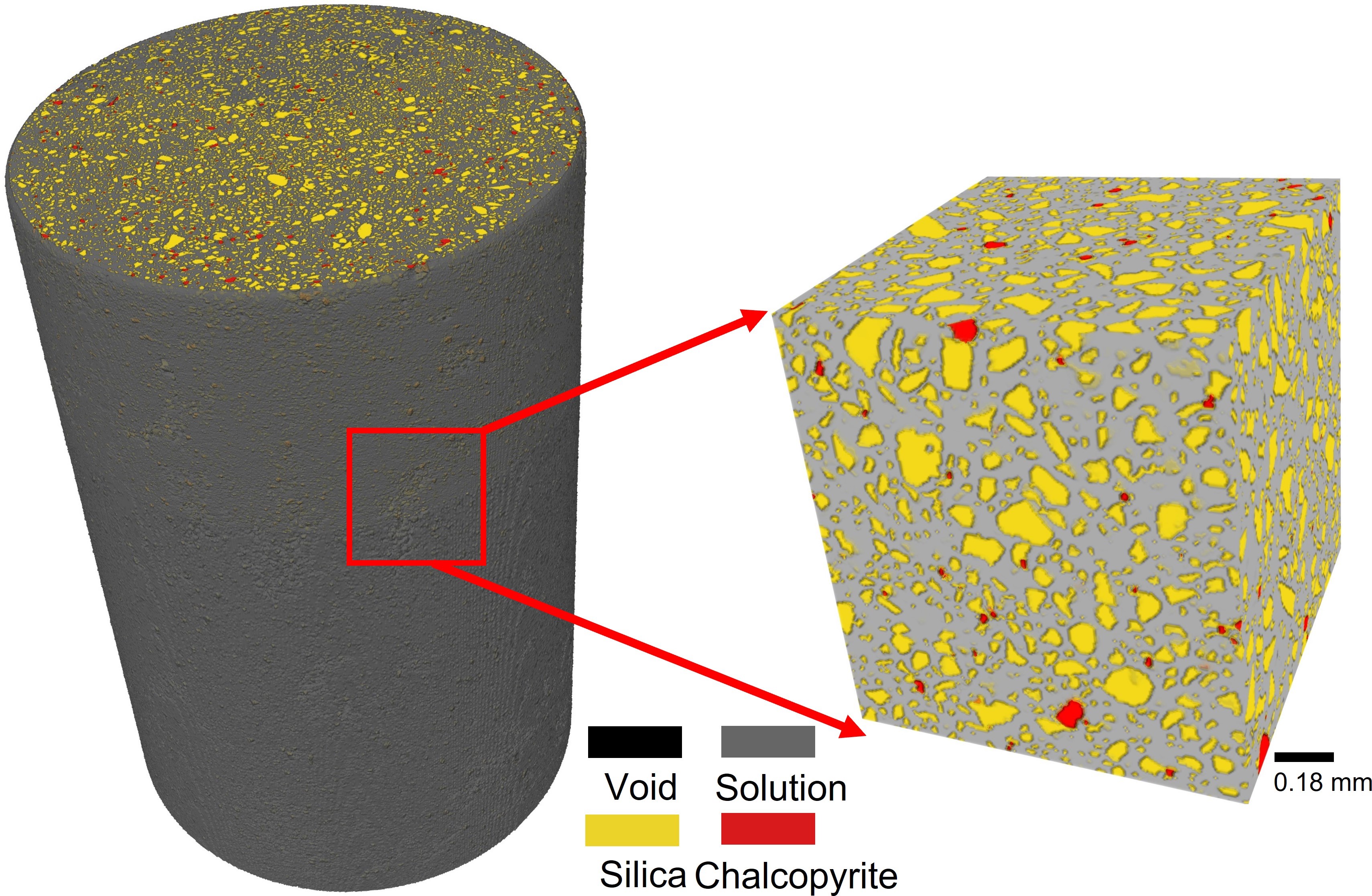}
    \caption{A representative element sub-domain with size of 256 cubic voxels. It was used for the EK transport simulation.}
    \label{fig:REV}
\end{figure}

\begin{figure}[H]
  \centering
    \includegraphics[width=\textwidth]{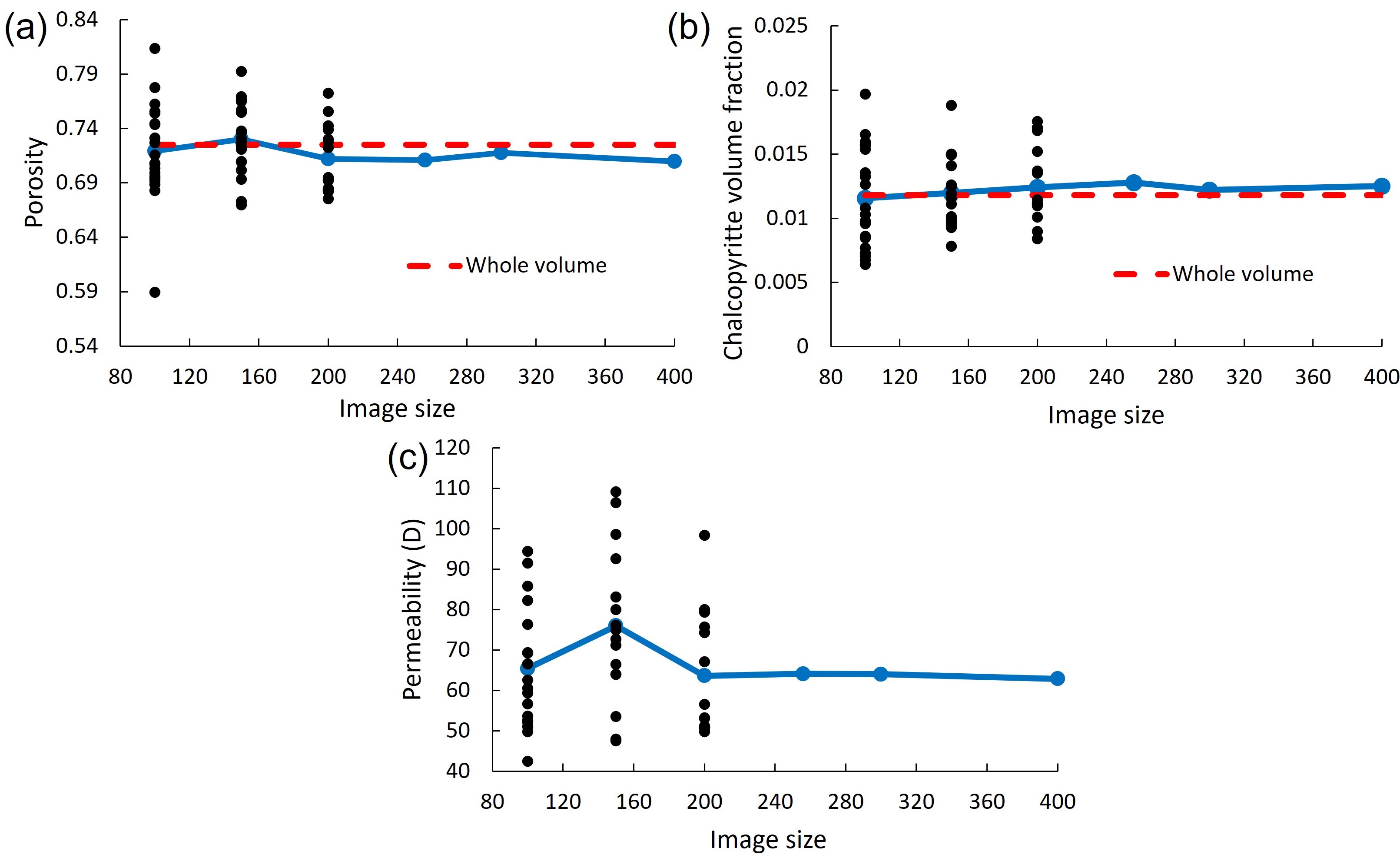}
    \caption{(a) Porosity, (b) Chalcopyrite volume fraction, (c) permeability as a function of image size (measured in number of voxels of the 3D image) for the synthetic ore system. Black points show the results for all sub-domains and the blue line is the average value of them.}
    \label{fig:homogeneous}
\end{figure}

A representative element sub-domain with size of 256 cubic voxels was cropped from the original domain, as shown in Fig. \ref{fig:REV}. To analyse the homogeneity of the synthetic ore system, several sub-domains with varied sizes were randomly cropped from the whole volume. Porosity, chalcopyrite volume fraction, and permeability were used to show the homogeneity of the system, as shown in Fig. \ref{fig:homogeneous}. The relationship of the image size to porosity and chalcopyrite volume fraction showed a linear relationship and the porosity values under different image sizes were closed to the value for whole volume, indicating that the ore synthetic is homogeneous. Moreover, the homogeneity for single phase EOF is comparable to the hydraulic single phase flow. Therefore, permeability of those sub-domains were also generated (Fig. \ref{fig:homogeneous} (c)). With image size less than $200^{3}$, the average permeability value showed a fluctuating trend, while with image size larger that $200^{3}$, the linear relationship can be observed. Therefore, the sub-domain size of $256^{3}$ is representative of the whole volume.

\subsection{Péclet number Calculation}
\label{sec:PE}

Péclet number can be calculated as

\begin{equation}\label{penumber}
    Pe = \frac{L\times \mu}{D},
\end{equation}

where $\mu$ is the average pore velocity, and $L$ is the characteristic length, calcualted by .

\begin{equation}\label{characteristic length}
    L = \frac{\pi \times V}{A},
\end{equation}

where $V$ is the bulk volume of the sample, and $A$ is the surface area of pores.

\renewcommand{\arraystretch}{1.6}
\begin{table}[H]
\centering
\caption{Peclet number calculation for Domain 1,2 and 3. EOF induced velocity is changed with chalcopyrite fraction, while electromigration induced velocity remains constant}
\begin{tabular}{lllllll}
\hline
Chalcopyrite   fraction & EOF induced velocity ($m/s$) & \multicolumn{4}{l}{Electromigration induced velocity ($m/s$) for H+ OH- Na+, Cl-, Fe3+}    \\
3.7\%                   & 1.40E-07 & \multicolumn{4}{l}{1.1E-05,  -6.1E-05,  1.5E-05,   -2.3E-05,     2.1E-05}\\
15.6\%                  & 1.14E-07 & \multicolumn{4}{l}{1.1E-05, -6.1E-05, 1.5E-05,  -2.3E-05,    2.1E-05}\\
63.3\%                  & 3.33E-08 & \multicolumn{4}{l}{1.1E-05, -6.1E-05, 1.5E-05,  -2.3E-05,    2.1E-05}\\ \hline
Chalcopyrite fraction   & $Pe$ (H$^+$)   & $Pe$ (OH$^-$)                      & $Pe$ (Na$^+$)   & $Pe$ (Cl$^-$)  & $Pe$ (Fe$^{3+}$)  &          \\ 
3.7\%                   & 0.994    & 0.991                        & 1.003     & 0.987    & 3.000     &          \\ 
15.6\%                  & 0.994    & 0.991                        & 1.001     & 0.988    & 2.996     &          \\ 
63.3\%                  & 0.993    & 0.993                        & 0.995     & 0.992    & 2.984     &    \\ \hline
\end{tabular}
\label{tab:PecletNumber}
\end{table}

\bibliographystyle{ieeetr}  
\bibliography{LBPM_COMSOL_paper}